\documentclass[journal]{IEEEtran}

\usepackage{amsmath} 
\usepackage{amssymb}  
\usepackage{array}
\usepackage[pdftex]{graphicx}
\usepackage{balance}
\usepackage{slashbox}
\usepackage{url}

	\usepackage[caption=false,font=footnotesize]{subfig}

\def\Ref#1{(\ref{#1})}

\newcommand{\x}{{\bf x}}
\newcommand{\bn}{{\bf n}}

\newcommand{\diag}{{\rm diag}}

\newcommand{\cN}{{\mathcal N}}

\newcommand{\cO}{{\mathcal O}}

\newcommand{\R}{\mathbb R}

\newcommand{\nn}{\nonumber}
\newcommand{\al}[1]{\begin{align}#1\end{align}}
\newcommand{\eq}[1]{\begin{equation}#1\end{equation}}

\newcommand{\erf}{{\rm erf}}
\newcommand{\alequal}{= & \,\:}
\newcommand{\aldef}{:= & \,\:}
\newcommand{\alleq}{\leq & \,\:}
\newcommand{\alnothing}{& \,\:}

\title{\LARGE \bf
A New Approach To Estimate The Collision Probability For Automotive Applications
}

\author{Richard Altendorfer
\thanks{Richard Altendorfer and Christoph Wilkmann are with Advanced Driver Assistance Systems,
ZF, Germany. {\tt\small \{Richard.Altendorfer, Christoph.Wilkmann\}@zf.com}}   
 and Christoph Wilkmann
}

\begin{document}

\maketitle
\thispagestyle{empty}
\pagestyle{empty}

\begin{abstract}
We revisit the computation of a probability of collision in the context of automotive collision avoidance (also referred to as conflict detection in other contexts).
After reviewing existing approaches to the definition and computation of a collision probability we 
argue that the question ``What is the probability of collision within the next three seconds?" can be answered on the basis of a collision probability rate.

Using results on level crossings for vector stochastic processes
we derive a general expression for the upper bound of the distribution of the collision probability rate. This expression is valid for arbitrary prediction models including process noise.

We demonstrate in several examples that distributions obtained by large-scale Monte-Carlo simulations obey this bound and in many cases approximately saturate the bound. 
We derive an approximation for the distribution of the collision probability rate that can be computed on an embedded platform.
An upper bound of the probability of collision is then obtained by one-dimensional numerical integration over the time period of interest.

A straightforward application of this method applies to the collision of an extended object with a second point-like object. Using an abstraction of the second object by salient points of its boundary we propose an application of this method to two extended objects with arbitrary orientation.

Finally, the distribution of the collision probability rate is compared to approximations of time-to-collision distributions for one-dimensional motions that have been obtained previously.
\end{abstract}
%
\section{Introduction}
The implementation of a collision mitigation or collision avoidance system requires the computation of a 
measure of criticality in order to assess the current traffic situation as well as its evolution in the short-term future. There are many criticality measures available, for example time-to-go (TTG) or time-to-collision (TTC) \cite{jansson2008framework},\cite{Muntzinger_et_al_09}, or 
the brake threat number \cite{stellet2015uncertainty}. All those measures are based on models of varying degrees of complexity of touching or penetrating the boundary of the potential colliding object, e. g. both the TTC $=-{x(0)\over \dot x(0)}$ (for a constant velocity model) and the brake threat number $a_{req}= -{\dot x^2(0)\over 2 x(0)}$ are based on the one-dimensional collision event $x(t) = 0$.

In this paper we focus on this underlying collision event -- the boundary penetration -- in a fully probabilistic manner, i. e. we propose a new approach to compute the collision probability for automotive applications.

The use of this collision probability for decision making in collision mitigation or avoidance systems is not subject of this investigation.

There are two different approaches to computing a collision probability for automotive applications that are known to the authors: 
\begin{enumerate}
\item probability of the spatial overlap of the host vehicle with the colliding vehicle's probability distribution, see \cite{jansson2005collision}, \cite{lambert2008fast}, and 
\item probability of penetrating a boundary around the host vehicle, see \cite{nordlund2008probabilistic}.
\end{enumerate}
There is currently no satisfying way to compute an automotive collision probability over a time period: there is a heuristic proposal to pick the maximal collision probability over that period as the collision probability for that time period \cite{jansson2008framework}, and there are calculations relying on strong assumptions (e. g. constant velocity models) that directly compute the collision probability over a time period \cite{nordlund2008probabilistic}.

On the other hand in the field of collision risk modeling for air traffic scenarios (for a recent overview see \cite{mitici2018mathematical}) a special case of the general mathematical result on crossings of multi-dimensional stochastic processes \cite{belyaev1968number} has been re-derived in \cite{bakker1993air} and applied to air traffic specific setups \cite{blom2002conflict},\cite{blom2003collision},\cite{cir319}. This allows for the computation of a collision probability over an extended period of time for aircraft modeled as axis-aligned cuboids or cylinders. Another approach based on a result for a one-dimensional stochastic process with particular dynamics has been suggested in \cite{prandini2000probabilistic}.

In the following, based on the formalism in \cite{belyaev1968number} we will derive an expression for the upper bound of the probability of penetrating a boundary around the host vehicle in a time period $\Delta T = [t_1, t_2]$. This will be the result of the temporal integration of an upper bound of the probability {\it rate} for which we derive a general expression valid for arbitrary prediction models including process noise. Inclusion of process noise is crucial for collision avoidance systems since it allows to encode the uncertainty in the relative motion of the host and the colliding vehicle. This uncertainty is particularly relevant in safety-critical applications with typical prediction times or TTCs of $\lesssim 5s$ where it is unknown whether the colliding vehicle keeps its motion, accelerates or slows down, or whether the host vehicle driver perceives the risk and slows down, for example.

The basis of our derivations are the time-dependent distributions $p_t( x, y, \dot x, \dot y, \dots), t\in\Delta T$. Those distributions characterize a non-stationary vector stochastic process that represents the predicted relative state $\xi^-(t)$ of the colliding vehicle. The stochastic process can be the result of a dynamical system whose flow $f$ can depend upon the state $\xi$, a time-dependent control input $u(t)$, process noise $\nu(t)$, and time $t$: 
\eq{
f\left( \xi, u(t), \nu(t), t \right)
}
In the remainder of this paper the time dependence of $\xi^-(t)$ and its elements will be suppressed, however the temporal dependence of probability distributions will be indicated by $p \rightarrow p_t$ where appropriate.

The expressions derived in this article are designed to be executed on embedded, automotive platforms; hence we do not resort to methods that include Monte-Carlo simulations as in \cite{blom2003collision} or \cite{jansson2008framework}. Instead we use large-scale Monte-Carlo simulations of potentially colliding trajectories as collision probability ground truth in order to assess the accuracy of our results. The Monte-Carlo outcome is represented by a histogram of the number of collisions that occur within a time interval with respect to time. Therefore simulating colliding trajectories naturally leads to the concept of a collision probability {\it rate} which is central to this article. 

The main contributions of this paper are as follows: 
\begin{enumerate}
\item the incorporation of the mathematical theory of level crossings of multi-dimensional stochastic processes developed in \cite{belyaev1968number} for the definition and computation of a general collision probability and derivation of upper bounds of the collision probability rate as well as the collision probability based on the entry intensity from \cite{belyaev1968number} (section \ref{sec_derivation_collision_prob})
\item derivation of approximate closed-form formulae for the collision probability rate for automotive applications (section \ref{sec_entry_intensity_approx} and app. \ref{app_computation_integral})
\item a novel numerical study of the collision probability rate with special emphasis on the accuracy of our approximate formulae as well as on the upper bound and its saturation (section \ref{section_numerical_study})
\item proposal of an adaptive method to efficiently sample the collision probability rate (section \ref{subsection_adaptive_method})
\item application of the computation of collision probability to a probabilistic treatment of two extended objects with arbitrary orientation by representative salient points of an object's geometry (sections \ref{sec_entry_intensity_extended_vehicles_theory} and \ref{sec_salient_point_entry_intensities})
\item comparison of the distribution of the collision probability rate to approximations of time-to-collision distributions for one-dimensional motions from existing approaches (section \ref{section_TTC})
\end{enumerate}
Note that while some of the contributions are specific to a two-dimensional setup and hence readily applicable to automotive problems others are generally applicable to collisions of higher-dimensional objects with piecewise smooth boundaries.

In the following two sections we will critically review existing approaches to computing a collision probability, first on the basis of spatial overlap and second with respect to boundary penetration.
\section{Collision probability from 2D spatial overlap}
\label{section_Collision_probability_from_2D_horizontal_integration}
This is the probability of the spatial overlap between the host vehicle and the colliding vehicle that has been investigated in \cite{blom2002conflict}, \cite{jansson2005collision}, \cite{lambert2008fast}. First, an instantaneous overlap probability is computed which involves integrals of the type
\eq{
P_{IO}(t) = \iiint_{(\psi, y, x)\in D} p_t\left( x, y, \psi \right) dx dy d\psi \label{eq_2D_integral}
}
where the state variables differ according to the model used (2D or 3D, with or without orientation angle $\psi$). The collision volume $D$ can be restricted to the vehicle boundary or can include a safe distance.
Note that even in the simplest case of only 2D position and Gaussian distribution the resulting two-dimensional integral, i. e. the cumulative distribution function of a bivariate Gaussian, cannot be solved in closed form; however, numerical approximation schemes exist \cite{genz2004numerical}. 
\begin{figure}[h]
\centering
\includegraphics[width = .9 \columnwidth]{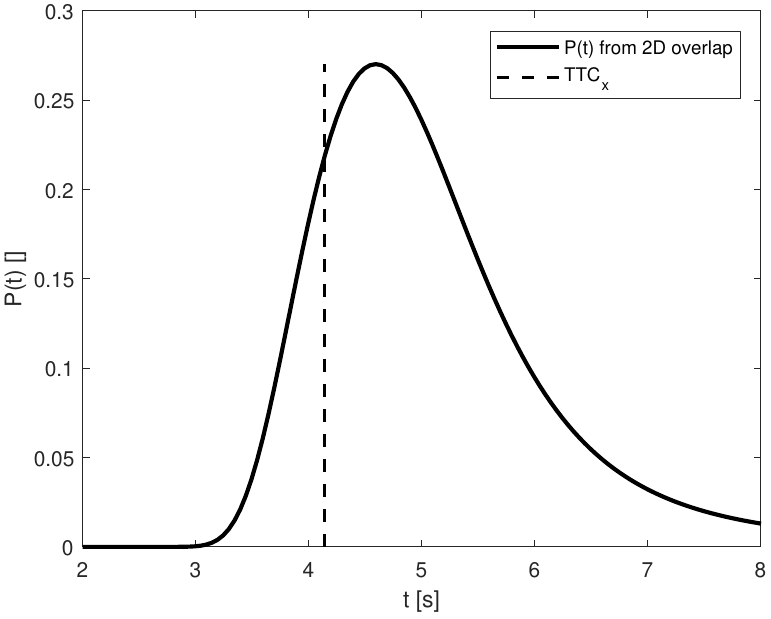}
\caption{Example of an instantaneous collision probability over time derived from a collision defined by spatial overlap as described in section \ref{section_Collision_probability_from_2D_horizontal_integration}. This is based on the first scenario described in sec. \ref{section_Collision_probability_Monte_Carlo_simulation} with initial condition in front of the host vehicle.}
\label{fig_collision_probability_spatial}
\end{figure}
A problem of deriving an instantaneous collision probability from 2D spatial overlap is that this approach directly yields a collision probability for a specific time, see fig. \ref{fig_collision_probability_spatial}. Hence it does not allow to answer the question ``What is the probability of collision within the next three seconds?" because integration of the collision probability over time does not yield a collision probability over a certain time period as already pointed out in \cite{nordlund2008probabilistic}.
In particular, time is not a random variable that can be marginalized over and an integral over a time interval $\Delta T$: $\int_{\Delta T}P_{IO}(t)dt$ has dimension of time and does not constitute a probability.
A heuristic proposal to solve this problem has been to
pick the maximal collision probability over a time period as the collision probability for that period \cite{prandini2000probabilistic},\cite{jansson2008framework}. This proposal has also been used in the definition of an overlap probability in \cite{blom2002conflict} where it was shown in a Monte-Carlo simulation that the overlap probability and a collision probability based on boundary crossings - discussed in the next section - are rather unrelated since they differ by two to three orders of magnitude. 

Another issue is that an instantaneous collision probability based on the overlap of a spatial probability distribution with the area of the host vehicle is determined by those sample trajectories whose current end points, i. e. the position at the current time, lie within the area of the host vehicle. But this is independent of when the trajectory has crossed the host vehicle boundary hence all end points except those exactly on the boundary (whose contribution to the two-dimensional integral is zero) correspond to a collision event in the past and therefore too late for collision avoidance, see also fig. \ref{fig_collision_probability_spatial} for an example where the maximum of the instantaneous collision probability from spatial overlap occurs after the TTC in x-direction.  
Also, by only considering trajectories with current end points within the area of the host vehicle other colliding trajectories with current end points outside the host vehicle area that have already entered and exited the boundary are unaccounted for. 
What we are actually interested in is the probability of the colliding object touching and/or penetrating the boundary of the host vehicle. 
This requires a different approach than integration over state space as in eq. \Ref{eq_2D_integral} since the integral over a lower-dimensional subspace would always be zero. Some existing approaches that consider a boundary instead of a state space volume for the computation of a collision probability are reviewed in the next section.
\section{Collision probability at boundary}
A probabilistic approach to computing the probability of penetrating a boundary - instead of the probability of a spatial overlap - has been proposed in \cite{nordlund2008probabilistic}. Their method is based on the probability density of the time to cross a straight, axis-aligned boundary assuming a constant velocity model. The derived collision probability refers to a time period and not just a time instant. It is only applicable to straight paths or combinations of piecewise straight paths and does not take into account more complex geometries such as a rectangle. It relies on a separation of longitudinal and lateral motion. Another limitation is that the stochastic nature of their conflict detection approach only comes from the distribution of the initial condition of their state -- process noise is not considered. 

A somewhat complementary approach is taken in \cite{prandini2000probabilistic} for aircraft conflict detection in the sense that process noise is incorporated whereas the uncertainty of the initial condition is not. They propose two different algorithms, one for mid-range and one for short-range conflict detection. For mid-range conflict detection their measure of criticality is an instantaneous probability of conflict and similar to the 2D spacial overlap discussed in the previous section. It is computed by a specific Monte-Carlo scheme. On the other hand their short-range conflict detection is based on the penetration of a spherical boundary around the aircraft as criticality measure. The dynamics is a constant velocity model perturbed by Brownian motion. The many strong assumptions, in particular constant velocity motion, specific Brownian noise model, and decoupling into one-dimensional motions make this approach hard to generalize.
 
The approaches discussed above are limited to constant velocity models with assumptions on the coupling of longitudinal and lateral motion, they either incorporate specific process noise or no process noise at all or exclude the uncertainty of the initial condition. Additionally, they all rely on a time-to-go or TTC as a prerequisite quantity - either probabilistic or non-probabilistic.

As we will show in the next section, such a temporal collision measure is not necessary for the computation of a collision probability.
Instead, we show that a fundamental quantity to compute the collision probability for stochastic processes is the collision probability {\it rate}. The mathematical foundation for this approach was provided in \cite{belyaev1968number} in terms of a stochastic intensity for piecewise smooth boundaries. Collision probability rates have already been used in the context of air traffic collision risk modeling: In \cite{bakker1993air} an intensity specific to a hyperrectangular boundary has been re-derived although \cite{belyaev1968number} was known to the authors. A subsequent publication (\cite{blom2003collision}) also includes definitions and theorems independent of established definitions and theorems from \cite{belyaev1968number} for those hyperrectangles.
Further applications to air traffic collision risk modeling are presented in \cite{blom2002conflict} and \cite{cir319}.
In the following section we define a collision probability and derive upper bounds of the collision probability and collision probability rate based on standard concepts from the theory of level crossings and
the intensity for more general boundaries from \cite{belyaev1968number}.
\section{Collision probability rate at boundary} 
\subsection{Derivation of an upper bound for the collision probability rate}
\label{sec_derivation_collision_prob}
We have seen that simulating colliding trajectories naturally gives us a probability rate and that a collision probability rate allows us to perform temporal integration to arrive at a collision probability for an extended period of time.
An expression for the upper bound of the collision probability rate will be derived on the basis of a theorem on boundary crossings of stochastic vector processes. For sake of lucidity of arguments we restrict ourselves to one of the four straight boundaries of the host vehicle, see fig. \ref{fig_front_boundary}; extension to the other boundaries is straightforward.

We start with the prediction of the pdf of a state vector that at least contains relative position and its derivative, i. e. $\xi = ( x\  y\  \dot x\  \dot y\  \cdots)^\top$ for a two-dimensional geometry, of a colliding object from an initial condition at $t=0$ to a future time $t$ where process noise $\nu(t)$ is explicitly incorporated:
\eq{
{\rm prediction:}\ p_0( x, y, \dot x, \dot y, \dots) \stackrel{t, \nu(t)}{\longmapsto} p_t( x, y, \dot x, \dot y, \dots) \nn
}
Note that we do not make any assumptions on the used prediction model as well as noise model or explicit temporal dependencies, hence the stochastic dynamical system that gives rise to this pdf could also explicitly depend upon time or a time-dependent control input $u(t)$.
In order to cast the following expressions into a more readable format we define a probability distribution that only depends upon relative position and its derivative by marginalization (see app. \ref{app_Partitioned_Gaussian_densities} for marginalization of Gaussian densities, for example) of the predicted pdf over the other variables:\footnote{The state vector $\xi = ( x\  y\  \dot x\  \dot y\  \ddot x\  \ddot y)^\top$ specified in app. \ref{app_vehicleModel}
is an obvious extension of the minimal state vector above with corresponding white noise jerk model described in eq. \Ref{eq_diff_equation} and is used as an example to illustrate the computation of collision probability rate. It is however by no means specific to the results stated in this paper.}
\eq{
p_t( x, y, \dot x, \dot y ) := \!\!\!\!\!\!\!\!\int\limits_{\rm other\ var.}\!\!\!\!\!\!\!\!p_t( x, y, \dot x, \dot y, {\rm other\ var.}) d({\rm other\ var.}) \nn
}
Given the pdf $p_t( x, y, \dot x, \dot y )$ what we are looking for is an expression for 
\eq{
{dP_C^+ \over dt }( \Gamma_{front}, t ) \nn
}
i. e. the collision probability rate ${dP_C^+ \over dt }$ with dimension $[{s}^{-1}]$ at time $t$ for the front boundary $\Gamma_{front}$. The superscript $+$ is used to denote that this probability rate is referring to boundary crossings from outside to inside.
\subsubsection{An Intuitive Motivation} \label{subsub_motivation}
We start with the probability of the colliding object being inside an infinitesimally thin strip at the boundary $\Gamma_{front}$ (see fig. \ref{fig_front_boundary})
\eq{
dP_C^+( \Gamma_{front}, t ) = 
\int\limits_{\dot y \in \R}\int\limits_{\dot x \leq 0}\int\limits_{y \in I_y} p_t( x_0, y, \dot x, \dot y )\, dx dy d\dot x d\dot y \nn
}
Here, since we are only interested in colliding trajectories, i. e. trajectories that cross the boundary from outside to inside, we do not fully marginalize over $\dot x$ but restrict the $x-$velocity to negative values at the boundary.

A collision probability rate can now be obtained by
dividing the unintegrated differential $dx$ by $dt$; in that way the ``flow" of the target vehicle through the host vehicle boundary is described at $x_0$ with velocity $\dot x \leq 0$:
\eq{
{dP_C^+ \over dt }( \Gamma_{front}, t ) \simeq -\!\!\!\!\!\int\limits_{\dot y \in \R}\int\limits_{\dot x \leq 0}\int\limits_{y \in I_y} p_t( x_0, y, \dot x, \dot y )\, \dot x\,  dy d\dot x d\dot y\label{eq_intuitive_collision_probability_rate}
}
Here, since the velocity is restricted to negative values a minus sign is required to obtain a positive rate.
This expression constitutes the expected value of the velocity component perpendicular to the boundary evaluated at the boundary.
\subsubsection{Derivation based upon the theory of level crossings}
This intuitive derivation can be amended as well as generalized in a mathematically rigorous way by invoking a result on crossings of a surface element by a stochastic vector process stated in \cite{belyaev1968number} and generalized in \cite{belyaev1969characteristics} and \cite{lindgren1980model}. First we need to set up the notations and definitions for entries and exits (level crossings) across the boundary of a region.

Let $\zeta(t)$ be a continuously differentiable $n-$dimensional vector stochastic process with values $\x \in \R^n$. The probability densities $p_t(\x)$ and $p_t(\bf{\dot x},\bf{x})$ exist where $\bf{\dot x} \in \R^n$ are the values of $\dot\zeta(t)$.\footnote{Further assumptions on the stochastic process and its probability densities apply \cite{belyaev1968number}.} Let the region $S\in \R^n$ be bounded by the smooth surface $\partial S$ defined by the smooth function $g$ as $\partial S = \{\x: g(\x) = 0\}$ and let $\Gamma \subseteq \partial S$ be a subset of that surface. Let $\bn_\Gamma(\x)$ be the surface normal at $\x$ directed towards the interior of the region.

A sample function $\x(t)$ of $\zeta(t)$ has an entry (exit) across the boundary $\Gamma$ at $t_0$ if $g(\x) > 0\ (g(\x) < 0) \forall t \in (t_0 - \epsilon, t_0)$ and $g(\x) < 0\ (g(\x) > 0) \forall t \in (t_0 , t_0 + \epsilon)$ for some $\epsilon > 0$. For a temporal interval $\Delta T = [t_1,t_2]$ the number of entries/exits across $\Gamma$ in this interval is denoted by $N^\pm(\Gamma, t_1, t_2 )$.

The importance of this mathematical setup is that
using the number of entries a collision probability over $\Delta T$ can be defined\footnote{This definition is motivated by the probability distribution of the maximum of a continuous process, see e. g. \cite{lindgren2012stationary}.} as
\al{
P^+_C( \Gamma, t_1, t_2 ) \aldef  P\left( g\left(x(t_1)\right) \geq 0, N^+(\Gamma, t_1, t_2 ) \geq 1 \right) \nn\\
\alnothing + \underbrace{P\left( g\left(x(t_1)\right) < 0\right)}_{=0} \nn\\
\alequal P\left( N^+(\Gamma, t_1, t_2 ) \geq 1\right) \label{eq_coll_prob_def_upper_bound}
}
i. e. the probability that the stochastic process enters the boundary in $\Delta T$ at least once with initial value outside the boundary.
The probability that the process is outside the boundary at initial time $t_1$ should be one
in automotive applications where a collision probability is to be computed for a time interval that begins at a time when the collision has not happened yet.

The first moment of $N^+(\Gamma, t_1, t_2 )$ can be used to obtain an upper bound for $P( N^+(\Gamma, t_1, t_2 ) \geq 1 )$:\footnote{Using Markov's generalized inequality also a lower bound can be derived in terms of the first and second factorial moments \cite{belyaev1968number}.}
\eq{
P( N^+(\Gamma, t_1, t_2 ) \geq 1 ) \leq {\rm E}\left\{ N^+(\Gamma, t_1, t_2 ) \right\} \label{eq_upper_bound}
}
This becomes obvious by writing out the expressions above: 
\eq{
P( N^+\!\!\geq 1 ) = \sum_{k=1}^\infty P( N^+\!\!= k ) \leq \sum_{k=0}^\infty k P( N^+\!\!= k ) = {\rm E}\left\{ N^+\!\right\} \label{eq_expected_number_inequality}
} 
It also shows that if the probabilities for two or more entries are much smaller than for one entry then ${\rm E}\left\{ N^+(\Gamma, t_1, t_2 ) \right\}$ is not just an upper bound but a good approximation to $P( N^+(\Gamma, t_1, t_2 ) \geq 1 )$.

It remains to compute the first moments for entry and exit which can be obtained via temporal integration of the entry/exit intensities $\mu^\pm$ as defined (see e. g. \cite{lindgren2012stationary}) below:
\eq{
\int\limits_{t_1}^{t_2} \mu^\pm (\Gamma, t ) dt := {\rm E}\left\{ N^\pm(\Gamma, t_1, t_2 ) \right\}
}
By combining eqs. \Ref{eq_coll_prob_def_upper_bound} and \Ref{eq_upper_bound} and evaluating the temporal derivative with respect to $t_2$ at $t_1$ we obtain
\eq{
{dP^+_C\over dt}\left( \Gamma, t_1 \right)  \leq \mu^+ (\Gamma, t_1 ) \label{eq_collision_probability_rate_upper_bound} 
}
i. e. we have derived an upper bound for the collision probability {\it rate}.

This upper bound can be further evaluated using the explicit expression for the entry/exit intensities $\mu^\pm$ from \cite{belyaev1968number}:
\eq{
\mu^\pm\left( \Gamma, t \right) =\!\!\!\int\limits_{\x\in\Gamma} {\rm E}\left\{\left. \langle\bn_\Gamma(\x),\dot\zeta(t)\rangle^\pm\right|\zeta(t) = \x\right\}p_t(\x)ds_\Gamma(\x) \label{eq_entry_exit_intensity}
}
where $\langle\cdot, \cdot \rangle$ is the scalar product, $ds_\Gamma(\x)$ is an infinitesimal surface element of $\Gamma$ at $\x$ and $(\cdot)^+ := \max(\cdot,0)$ and $(\cdot)^- := -\min(\cdot,0)$. Equation \Ref{eq_entry_exit_intensity} holds for general non-Gaussian as well as non-stationary stochastic processes.

\begin{figure}[h]
\centering
\includegraphics[viewport = 7cm 3cm 20cm 14.2cm, clip, width = 0.7 \columnwidth]{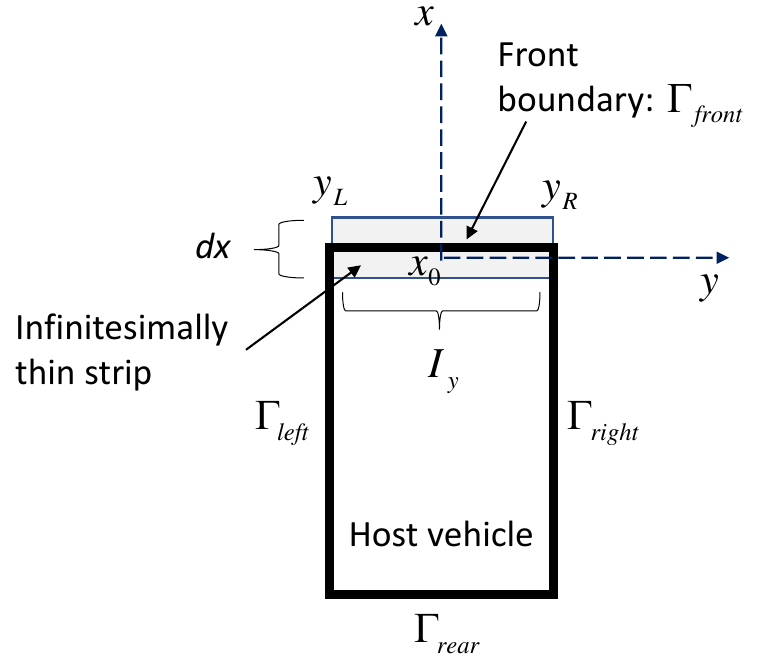}
\caption{Horizontal view of the host vehicle rectangle with local Cartesian coordinate system and coordinate origin at the
middle of the front boundary characterized by $x=0$ and $y \in [y_L, y_R] = I_y$.}
\label{fig_front_boundary}
\end{figure}

In order to apply eq. \Ref{eq_entry_exit_intensity} to the front boundary $\Gamma_{front}$ as in fig. \ref{fig_front_boundary} we need to perform the following identifications:\footnote{From now on we now do not distinguish anymore between a stochastic process and its sample values.}
\al{
\zeta(t) \alequal \left( x, y \right)^\top \nn\\
\Gamma_{front} \alequal \{(x, y): x - x_0 = 0 \wedge y\in I_y \} \nn\\
g_{\Gamma_{front}}(\x) \alequal x - x_0 \nn\\
\bn_{\Gamma_{front}}(\x) \alequal \left( -1, 0 \right)^\top \nn\\
ds_{\Gamma_{front}}(\x) \alequal dy
}
Hence we obtain for the intermediate expectation operator
\begin{multline}
{\rm E}\left\{\left. \langle\bn_{\Gamma_{front}}(\x),\dot\zeta(t)\rangle^+\right|\zeta(t) = \x\right\} =\\
-\int\limits_{\dot y \in \R}\int\limits_{\dot x \leq 0} \dot x\, p_t( \dot x, \dot y | x, y )\, d\dot x d\dot y
\end{multline}
and the entry intensity becomes
\begin{align}
\!\mu^+\!\left( \Gamma_{front}, t \right)&\!=\!-\!\!\!\!\!\int\limits_{y \in I_y}\!\!\!\!\!\left(\, \int\limits_{\dot y \in \R}\int\limits_{\dot x \leq 0}\!\!\! \dot x\, p_t( \dot x, \dot y | x_0, y ) d\dot x d\dot y\!\!\right)\!\! p_t( x_0, y ) dy \nn\\
&=\!-\!\!\!\!\!\int\limits_{\dot y \in \R}\int\limits_{\dot x \leq 0}\int\limits_{y \in I_y} \dot x\, p_t( x_0, y, \dot x, \dot y )\, dy d\dot x d\dot y\label{eq_front_boundary_entry_intensity}
\end{align}
This shows that the intuitive derivation of the collision probability rate (eq. \Ref{eq_intuitive_collision_probability_rate})
results in the correct expression for the upper bound.
It should be noted, however, that the application of the formalism above to a rectangular boundary of the host vehicle is just an example. By the theorem stated above the formula can be applied to any subsets of smooth surfaces, including higher dimensional ones for three-dimensional objects, for example.

The computation above applies to the front boundary of the host vehicle. Since the results in \cite{belyaev1968number} are also valid for piecewise smooth boundaries\footnote{This corrects a contrary statement in \cite{mitici2018mathematical}. The results of \cite{belyaev1968number} have been applied to polyhedral \cite{veneziano1977vector} and other regions $S$ with a piecewise smooth surface $\partial S$, for an overview see \cite{illsley1998moments}.} the entry intensities of the four boundaries can be added. Hence the total entry intensity is given by
\al{
\!\!\!\!\!\!\!\mu^+(\Gamma_{host\ vehicle}, t ) \alequal \mu^+(\Gamma_{front}, t ) + \mu^+(\Gamma_{right}, t )\nn\\
 \alnothing + \mu^+(\Gamma_{left}, t ) + \mu^+(\Gamma_{rear}, t )
}
With these expressions the collision probability rate and collision probability for the surface subset $\Gamma$ within a time interval $\Delta T = [t_1,t_2]$ are bounded by
\al{
\!\!\!\!\!\!\! {dP^+_C\over dt}\left( \Gamma_{host\ vehicle}, t_1 \right)  \alleq \mu^+ (\Gamma_{host\ vehicle}, t_1 ) \\
\!\!\!\!\!\!\! P^+_C( \Gamma_{host\ vehicle}, t_1, t_2 ) \alleq \int\limits_{t_1}^{t_2} \mu^+\left( \Gamma_{host\ vehicle}, t \right) dt \label{eq_temporal_integration}
}
In summary the upper bounds are due to the approximation of the probability of one or more boundary entries by the expected number of boundary entries (inequality \Ref{eq_upper_bound}).

Note that the stochastic process $\xi$ representing the state of the colliding object needs to contain 2D relative position $(x\ y)^\top$ and 2D relative velocity $(\dot x\ \dot y)^\top$. In many ADAS applications the target vehicle dynamics for moving as well as stationary vehicles is modeled directly in relative coordinates. For state vectors that do not contain the 2D relative velocity but other quantities such as the velocity over ground (see e. g. \cite{Altendorfer09}), a probabilistic transformation to relative velocities must be performed first. 
\subsection{Entry intensity for two extended vehicles}
\label{sec_entry_intensity_extended_vehicles_theory}
\begin{figure}[h]
\centering
\includegraphics[viewport = 4cm 4.1cm 24cm 15.3cm, clip, width = 0.5 \columnwidth]{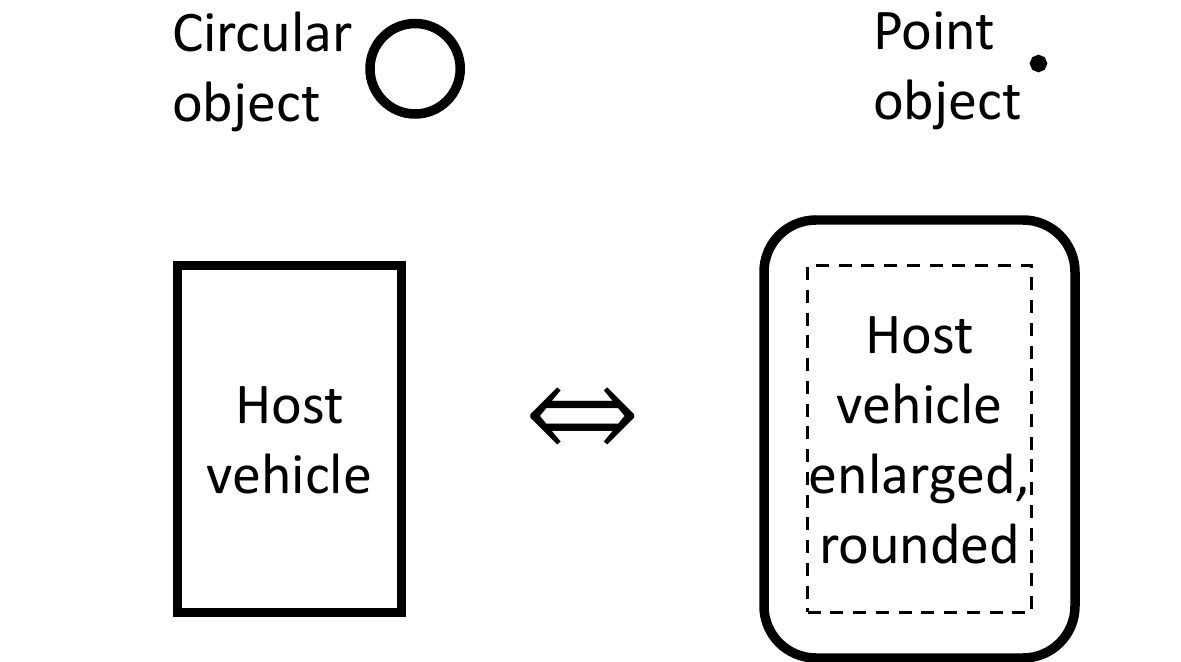}
\caption{Equivalent representations of objects regarding collision modeling.}
\label{fig_boundary_equivalence}
\end{figure}

In previous sections the colliding vehicle was modeled as a point distribution corresponding to a single reference point (for example the middle of the vehicle's rear bumper). This allowed the direct application of the theory of boundary crossings of a point process. Using strong assumptions about the two colliding objects' shape and orientation this can also be applied to two extended objects: the collision models in \cite{cir319} assume either a partially isotropic shape (cylinder) or use axis-aligned cuboids and can hence be reduced to the collision of an extended object and a point distribution as described in \cite{cir319}. Note that the collision of a rectangular host vehicle with a circular object (this could serve as the approximate horizontal shape of a pedestrian) can also be reduced to the collision of the circle midpoint with an enlarged ``rectangle" with rounded edges as shown in fig. \ref{fig_boundary_equivalence}.
\newline
While in many ADAS applications the object state is modeled by a single reference point and possibly additional attributes, for many collision avoidance applications the object extension/geometry is crucial, e. g. for scenarios with fractional overlap. 
In \cite{philipp2019analytic} a method to compute the collision probability between a rectangular host and rectangular object vehicle was derived using our method described above. By assuming both the host trajectory and the object orientation to be deterministic the collision between two oriented rectangles was reduced to the collision of a point and an octagon. 
\newline
However, for example in EuroNCAP-relevant junction scenarios where an oncoming object turns into the host vehicle's path the latter assumption might not be valid.
Instead, in order to represent the extended geometry of the colliding object, we model the object's dynamical state (see app. \ref{app_vehicleModel}) relative to a specific reference point, for example the middle of the rear bumper or the rear axle, and then transform the state probability distribution to its boundary (see app. \ref{app_state_transformation_salient}).
\begin{figure}[h]
\centering
\includegraphics[viewport = 9cm 2cm 21cm 18cm, clip, width = 0.5 \columnwidth]{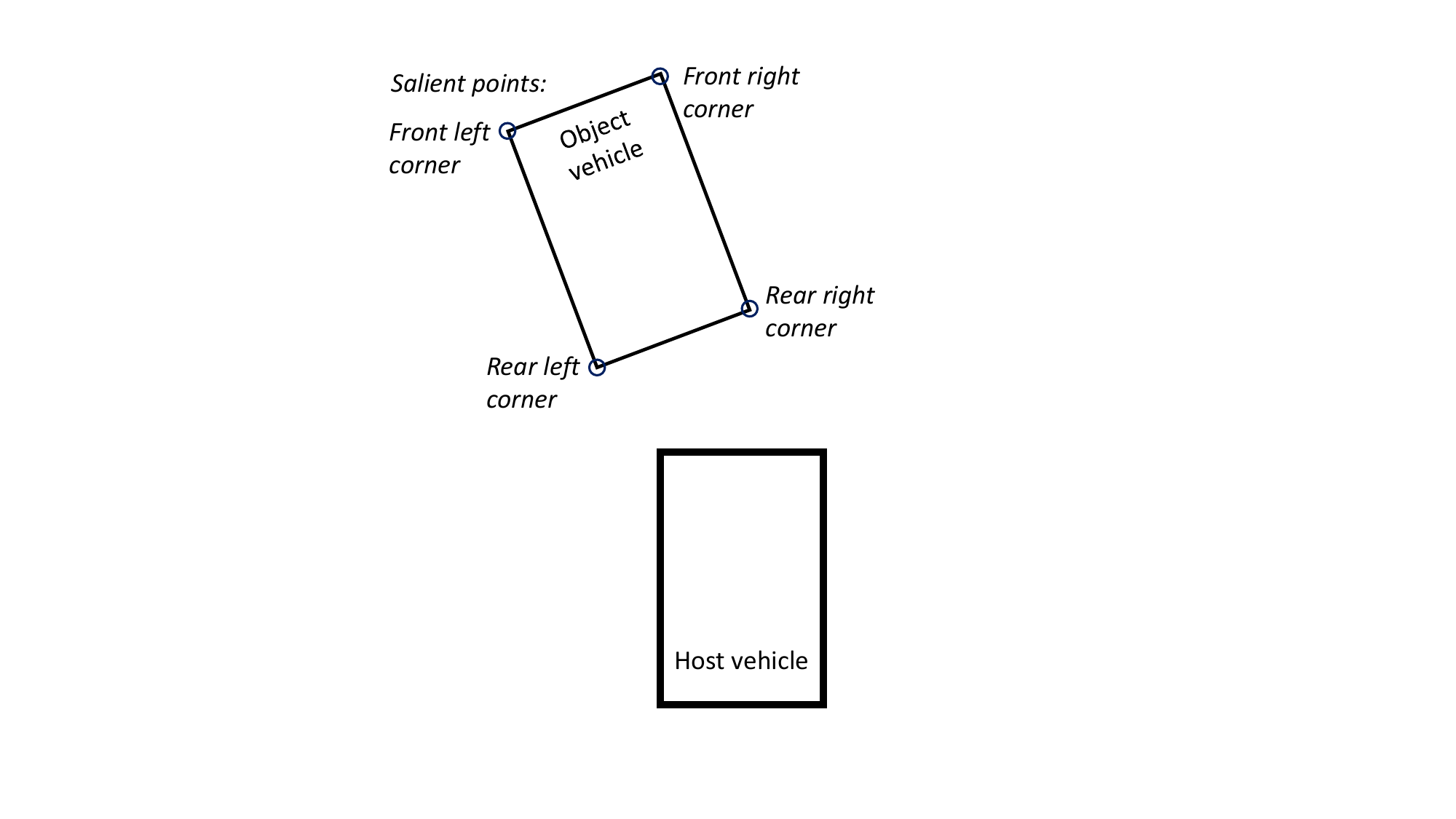}
\caption{Horizontal view of the host vehicle and colliding object vehicle with reference point and salient points at the four corners.}
\label{fig_host_and_object_vehicle}
\end{figure}
The entry intensity can be obtained for every point on the colliding object's boundary which results in a family of entry intensities. We approximate this family by the entry intensities of a small number of representative salient points of the colliding vehicle's two-dimensional geometry as in fig. \ref{fig_host_and_object_vehicle}. The collision probability for the extended colliding object can then be approximately determined by the collision probability of the ``riskiest" salient point which we define to be the one where the collision probability exceeds a certain threshold the earliest. A fully worked example of this approach is given in sec. \ref{sec_salient_point_entry_intensities}.
\subsection{Implementation for Gaussian distributions}
\label{sec_entry_intensity_approx}
For further computations - especially in the Gaussian case - it will be convenient to marginalize over $\dot y$ and rewrite eq. \Ref{eq_front_boundary_entry_intensity} in terms of a conditional probability:
\eq{
\mu^+( \Gamma_{front}, t ) = -p_t( x_0 ) \int\limits_{y \in I_y} \int\limits_{\dot x \leq 0} \dot x\, p_t( \dot x, y | x_0 )\, d\dot x dy \label{eq_front_boundary_entry_intensity_cond}
}
For general distribution functions the integral in eq. \Ref{eq_front_boundary_entry_intensity_cond} cannot be computed in closed form and numerical integration methods must be used. Even in the bivariate Gaussian case there is no explicit solution known to the authors. However, by a Taylor-expansion with respect to the off-diagonal element of the inverse covariance matrix of $p( y, \dot x | x_0 )$ as detailed in app. \ref{app_computation_integral}, the integral can be factorized into one-dimensional Gaussians and solved in terms of the standard normal one-dimensional cumulative distribution function $\Phi$. To zeroth order the integration yields:
\begin{align}
\mu^+( \Gamma_{front}, t )=&-\cN( x_0; \mu_x, \sigma_{x} ) \nn\\
& \cdot \Bigg(\bigg( \mu_{\dot x|x_0} \Phi\left({ - \mu_{\dot x|x_0} \over \tilde\sigma_{\dot x|x_0}}\right) \nn \\
&  \quad -\tilde\sigma_{\dot x|x_0}^2 \cN( 0; \mu_{\dot x|x_0}, \tilde\sigma_{\dot x|x_0} )  \bigg)\cdot \nn \\
& \cdot \left( \Phi\left({ y_R - \mu_{y|x_0} \over \tilde\sigma_{y|x_0}}\right) - \Phi\left({ y_L - \mu_{y|x_0} \over \tilde\sigma_{y|x_0}}\right) \right) \nn\\
& + \cO\left(\Sigma^{-1}_{12}\right) \Bigg)\label{eq_coll_prob_rate_num_approx}
\end{align}
Here, if $\Sigma\in \R^{2\times 2}$ is the covariance matrix of $p( \dot x, y | x_0 )$, then $\tilde\sigma_{\dot x|x_0} = \sqrt{|\Sigma|\over \Sigma_{yy}}$ and $\tilde\sigma_{y|x_0} = \sqrt{|\Sigma|\over \Sigma_{\dot x\dot x}}$, see app. \ref{app_computation_integral} where the integration has also been carried out to first order in $\Sigma^{-1}_{12}$.
Expression \Ref{eq_coll_prob_rate_num_approx} can be computed on an embedded platform using 
the complementary error function available in the C math library.\footnote{$\Phi$ is related to the error function ${\rm erf}$ and complementary error function ${\rm erfc}$ by $\Phi(x) = {1 \over 2}{\rm erfc}\left( {-x \over \sqrt{2}}  \right) = {1 \over 2} - {1 \over 2}{\rm erf}\left( {-x \over \sqrt{2}}  \right) $.}

In the next section an extensive numerical study using the above formulae and Monte-Carlo simulations is presented.
\section{Numerical Study}
\label{section_numerical_study}
\subsection{The collision probability ground truth: large-scale Monte-Carlo simulations}
\label{section_Collision_probability_Monte_Carlo_simulation}
Here, we want to investigate two examples of possible collision scenarios, one where the target vehicle is currently in front of the host vehicle and one where it is on the front right side. 
In order to obtain ground truth data for the future collision probability Monte-Carlo simulations are performed. The target vehicle on a possibly colliding path with the host vehicle is modeled by the state vector $\xi = ( x\  y\  \dot x\  \dot y\  \ddot x\  \ddot y)^\top$ and the dynamical system as specified in appendix \ref{app_vehicleModel}. The target vehicle is chosen to be detected by a radar sensor mounted at the middle of the front bumper of the host vehicle.
Note however that this state vector as well as the dynamical system specified in appendix \ref{app_vehicleModel} constitute just an example -- the central results in section \ref{sec_derivation_collision_prob} hold for general non-stationary as well as non-Gaussian stochastic processes. In particular, the absence of assumptions on the stationarity of the stochastic process means
that processes derived from more general dynamical system -- including systems with explicit time dependence or time-dependent control inputs $u(t)$ -- are covered.

\begin{figure*}[ht]
\centering
\null\hfill
\subfloat[The target is coming from the front $\left(\mu_x,\mu_y\right)=\left(10,0\right)m$. The parameters for the time-dependent input as specified in app. \ref{app_vehicleModel} are $b_1 = -0.2 m s^{-3}, b_2 = -0.3 m s^{-3}, \omega = 0.5 s^{-1}$.]{\includegraphics[width = .9 \columnwidth]{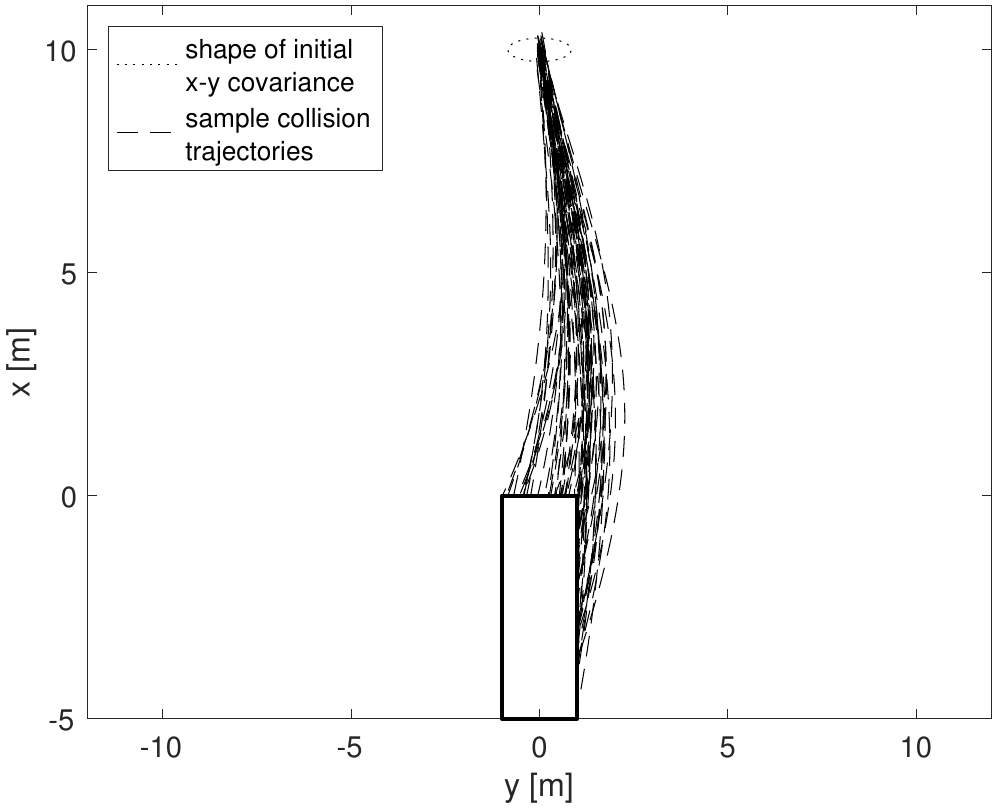}}
\hfill
\subfloat[The target is coming from the front right $\left(\mu_x,\mu_y\right)=\left(10,10\right)m$. The parameters for the time-dependent input as specified in app. \ref{app_vehicleModel} are $b_1 = -0.4 m s^{-3}, b_2 = -0.5 m s^{-3}, \omega = 0.5 s^{-1}$.]{\includegraphics[width = .9 \columnwidth]{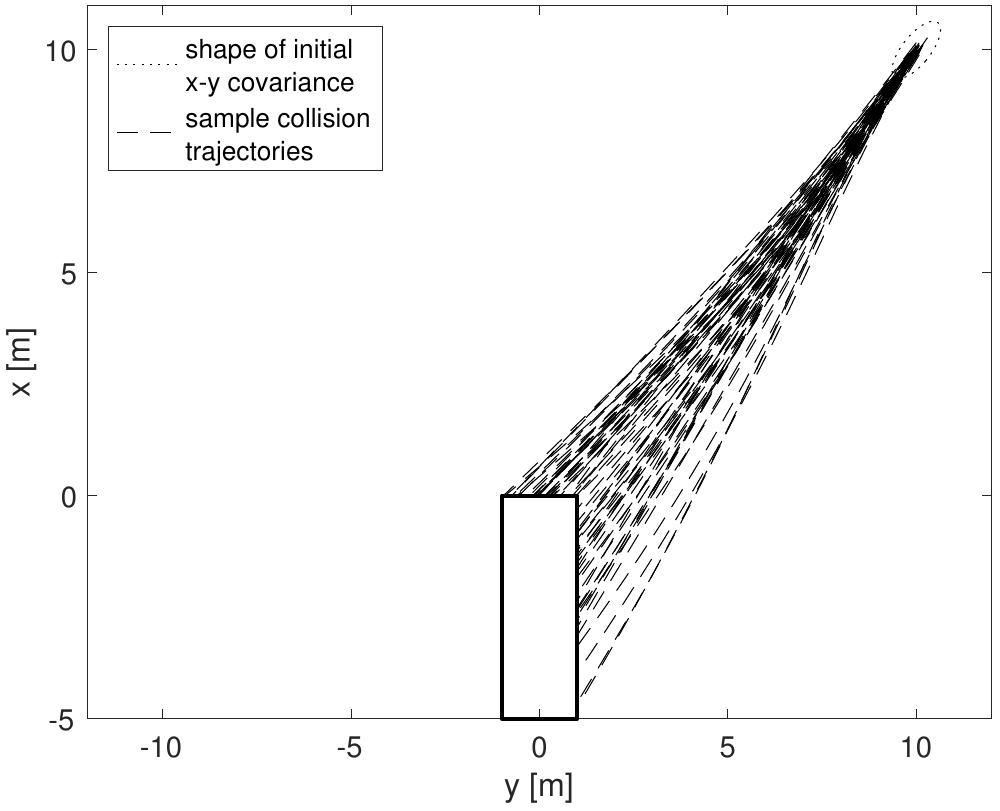}}
\hfill\null
\caption{Samples of simulated colliding trajectories for vehicles initially coming from the front (a) and from the front right (b) side.}
\label{fig_Monte_Carlo_sample_trajectory}
\end{figure*}

The starting point for an individual simulation is a sample point in state space $\xi_i^-$ where the target vehicle is some distance away from the host vehicle - either directly in front or coming from the right side, see fig. \ref{fig_Monte_Carlo_sample_trajectory}. This sample point is drawn from a multivariate distribution characterized by its mean vector and covariance matrix which is usually the output of a probabilistic filter that takes into account the history of all previous sensor measurements that have been associated with this object. Instead of arbitrarily picking specific values for this {\it initial} covariance matrix we take its values from 
steady state at this mean vector using the discrete algebraic Riccati equation for typical radar detection measurements.\footnote{Strictly speaking there is no steady state at those points since the system is non-linear and the relative speed is not zero. Nevertheless the solution of the Riccati equation is still representative if the filter settles within a smaller time period than the time period in which the state changes significantly.}
An instance $\xi_i^-$ of an initial state of the target vehicle is drawn as a sample of $\cN(\xi^-; \mu_{\xi}^-, P^-_{\infty})$. This state is predicted using the stochastic differential equation \Ref{eq_diff_equation} until it crosses the host vehicle boundary or a certain time limit is exceeded. 
Hence: collision event = crossing of the target vehicle path with the host vehicle boundary.
The time until the crossing is recorded and a new simulation with a new sample of initial conditions is started. Examples of colliding trajectories starting from an initial position in front of the host vehicle are depicted in fig. \ref{fig_Monte_Carlo_sample_trajectory}.
We have performed simulations of $N_{traj} = 3\cdot 10^6$ trajectories for the two starting points. The result is represented by a histogram of the number of collisions that occur within a histogram bin, i. e. time interval, with respect to time.

Hence simulating colliding trajectories naturally leads to a collision probability {\it rate} which is by construction the distribution of the TTC.

\begin{figure}[h]
\centering
\includegraphics[width = .9 \columnwidth]{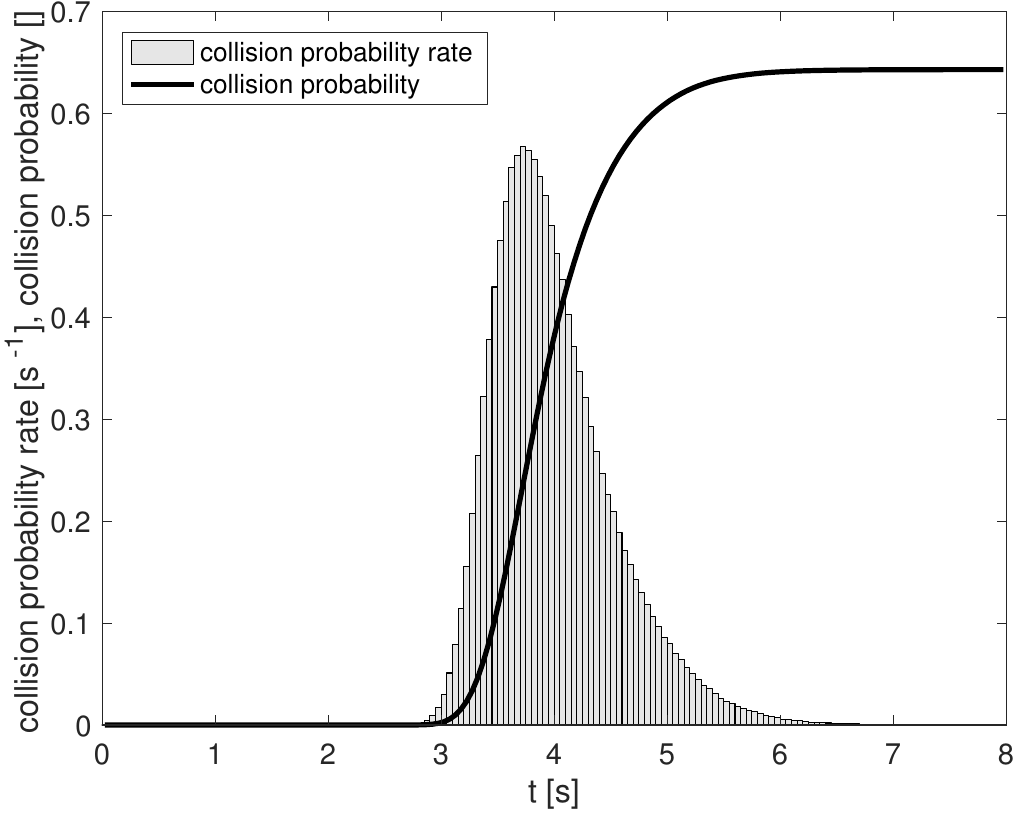}
\caption{Collision probability rate as a function of time for $\left(\mu_x,\mu_y\right)=\left(10,0\right)m$ based upon $N_{traj} = 3\cdot 10^6$ trajectories. Also shown is the collision probability obtained by integrating over time. This should be contrasted with the shape of the instantaneous collision probability in fig. \ref{fig_collision_probability_spatial}.}
\label{fig_Monte_Carlo_collision_probability_rate}
\end{figure}

An example is given in fig. \ref{fig_Monte_Carlo_collision_probability_rate} where the bins are normalized by the total number of trajectories $N_{traj}$ and the chosen bin width of $dt=0.05s$ to obtain a collision probability rate. In addition, the collision probability rate integrated by simple midpoint quadrature from 0 to time $t$ is shown. In this example the probability of collision with the target vehicle exceeds $60\%$ within the first $6s$. The asymptotic value of the collision probability as $t \rightarrow \infty$ indicates the overall probability of collision over all times.
With the setup explained above the following questions can be addressed:
\begin{itemize}
\item Is the expression for calculating the entry intensity from eq. \Ref{eq_entry_exit_intensity} consistent with the results from large scale Monte-Carlo simulations?
\item How does the approximation \Ref{eq_coll_prob_rate_num_approx} perform in comparison with the numerical integration of the derived expression \Ref{eq_front_boundary_entry_intensity_cond} for the entry intensity?
\item Can the computational effort be reduced by increasing $\Delta t$ and still accurately calculating the entry intensity?
\item Does the entry intensity still reproduce results from Monte-Carlo simulations after non-linear transformation from a reference point to representative salient points of the colliding vehicle's geometry? 
\end{itemize}
\subsection{Is the upper bound of the collision probability rate corroborated by Monte-Carlo simulation?}
\label{sec_numerical_corroboration}
\begin{figure*}[h!]
\centering
\null\hfill
\subfloat[Front scenario total collision probability rate and entry intensity]{\includegraphics[width = .7 \columnwidth]{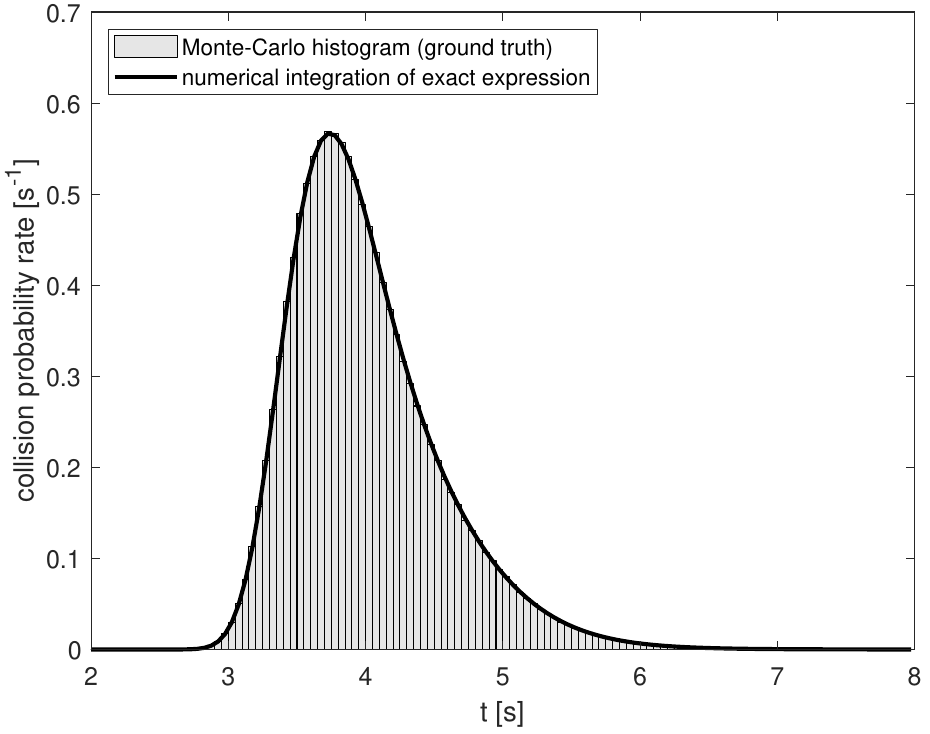}}
\hfill
\subfloat[Front-Right scenario total collision probability rate and entry intensity]{\includegraphics[width = .7 \columnwidth]{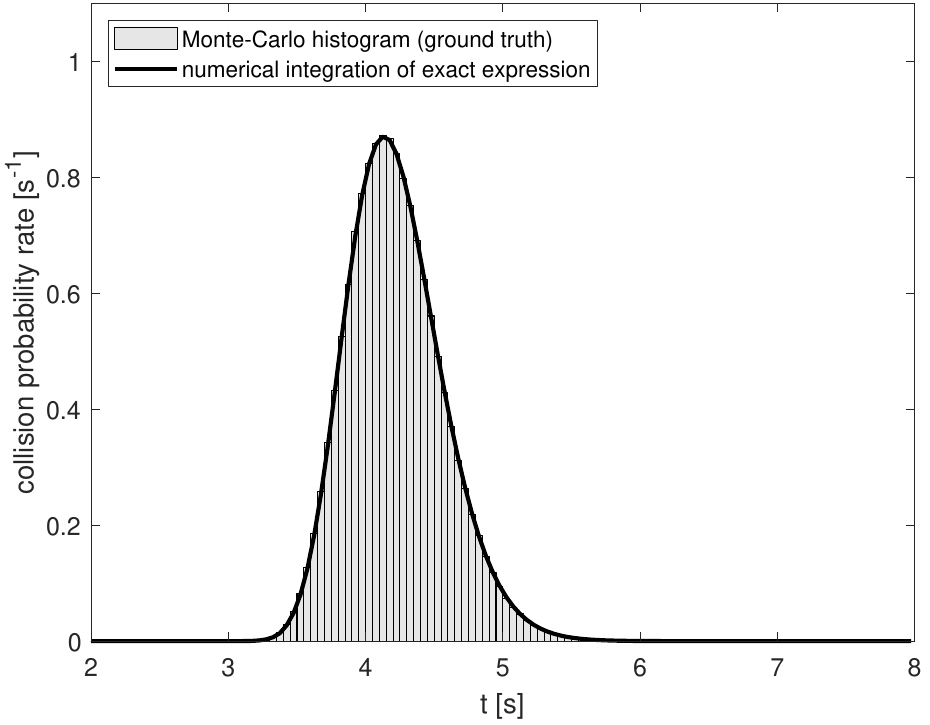}}
\hfill\null\\
\null\hfill
\subfloat[Front scenario: difference between total collision probability rate and entry intensity]{\includegraphics[width = .7 \columnwidth]{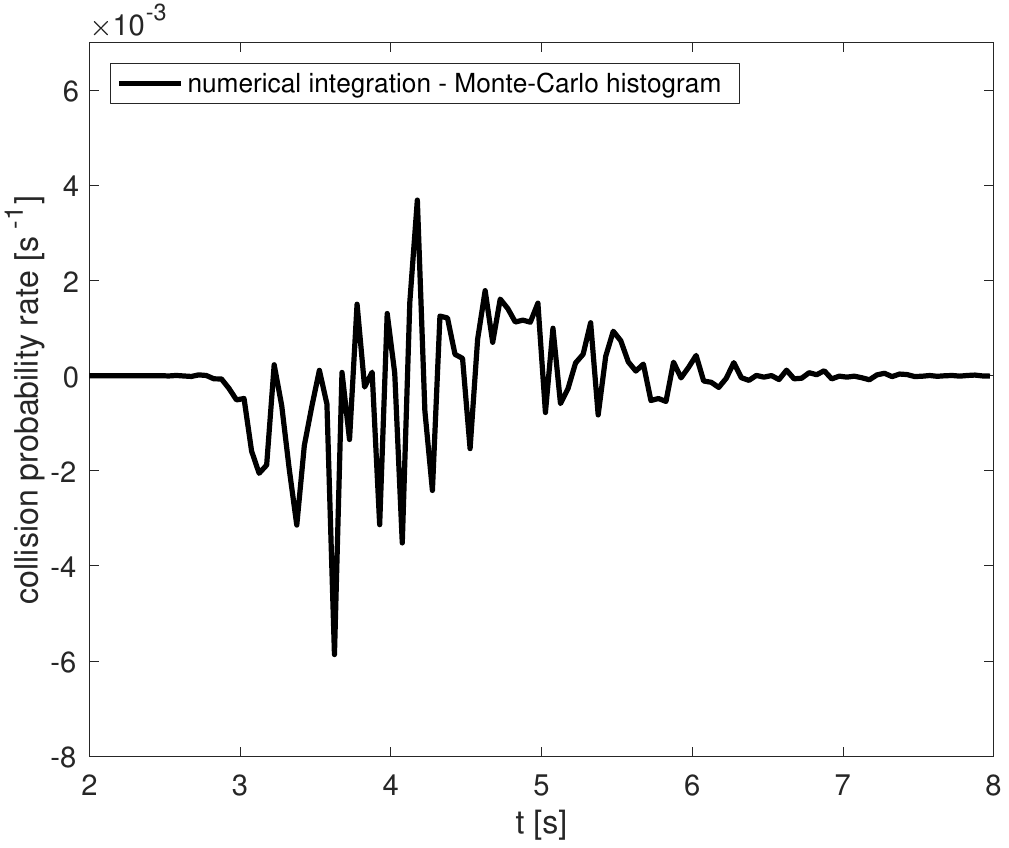}}
\hfill
\subfloat[Front-Right scenario: difference between total collision probability rate and entry intensity]{\includegraphics[width = .7 \columnwidth]{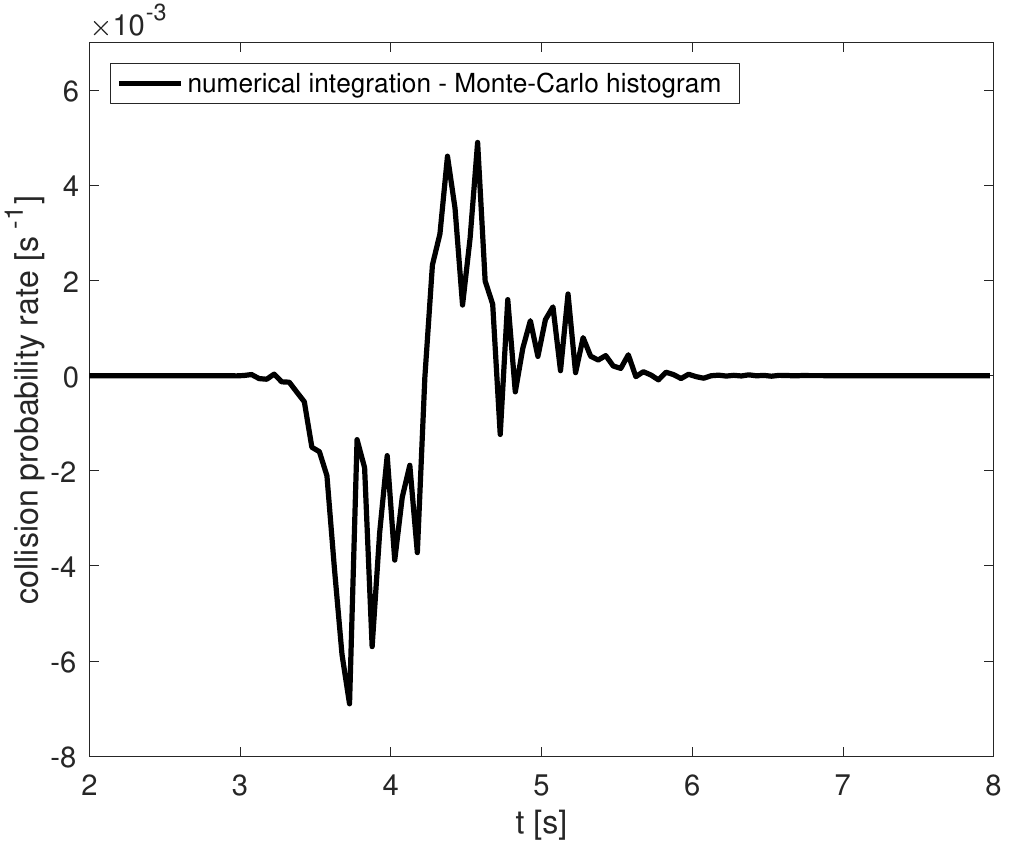}}
\hfill\null
\caption{The histogram resulting from Monte-Carlo simulation is shown together with the entry intensity obtained by numerical integration of the bivariate Gaussian for front (a) and front-right (b) scenario. The differences between simulation and numerical integration are calculated by evaluating the numerical integration at the same time as the mid points of the histogram bins and shown in (c) and (d). The process noise PSD for both coordinates is $\tilde q_x = \tilde q_y = 0.0101 m^2 s^{-5}$.}
\label{fig_Monte_Carlo_Histogram}
\end{figure*}

In order to address the first question, large scale Monte-Carlo simulations as described in sec. \ref{section_Collision_probability_Monte_Carlo_simulation} have been performed. Entry intensities were calculated based on $3\cdot 10^6$ sample trajectories for each of the two initial conditions $\cN_i(\xi^-; {\mu_{\xi}^-}_i, {P^-_{\infty}})$, where ${\mu_{\xi}^-}_i$ is shown in table \ref{tab_intial_conditions_for_simulation}, and ${P^-_{\infty}}$ is calculated using the discrete Riccati equation with matrices defined in appendix \ref{app_vehicleModel}. The two initial conditions $\left(i\in\left\{f, fr\right\}\right)$ describe a starting point directly in front of the host vehicle, and in front to the right at an angle of $45$ degrees with respect to the host vehicle. The inclusion of process noise as well as the inclusion of acceleration in the state vector as specified in appendix \ref{app_vehicleModel} allows for multiple entries and enables us to assess the influence of multiple entries on the accuracy of the upper bounds derived above.\footnote{This is in contrast to the aviation-specific numerical study in \cite{blom2003collision} where during the part of a trajectory where a collision could occur a constant velocity model without process noise was used and hence multiple entries were excluded.}
\begin{table}[h]
\caption{Mean of initial conditions for Monte-Carlo simulations}
\label{tab_intial_conditions_for_simulation}
\begin{center}
\begin{tabular}{ m{2cm} m{2cm} m{2cm} }
 \hline
${\mu_{\xi}^-}$ & Scenario & \\
\cline{2-3}
& front $(f)$ & front right $(fr)$ \\
  \hline
  ${\mu_{x}^-} \left[m\right]$ & $10$ & $10$ \\
  ${\mu_{y}^-} \left[m\right]$ & $0$ & $10$ \\
  ${\mu_{\dot x}^-} \left[\frac{m}{s}\right]$ & $-2$ & $-2$ \\
  ${\mu_{\dot y}^-} \left[\frac{m}{s}\right]$ & $0.4$ & $-1.6$ \\
  ${\mu_{\ddot x}^-} \left[\frac{m}{s^2}\right]$ & $-0.2$ & $-0.001$ \\
  ${\mu_{\ddot y}^-} \left[\frac{m}{s^2}\right]$ & $0.0$ & $-0.01$ \\
  \hline
\end{tabular}
\end{center}
\vspace{-2mm}
\end{table}
Table \ref{tab_number_collisions} shows the number of collisions divided into the respective boundaries of the host vehicle where the impact or boundary crossing occurred for the two different simulations. 
\begin{table}[h]
\caption{Number of collisions at host vehicle boundaries for $3\cdot10^6$ simulated trajectories with different initial conditions.}
\begin{center}
\label{tab_number_collisions}
\begin{tabular}{ m{2cm} m{2cm} m{2cm} }
 \hline
  $\Gamma$ & Scenario & \\
\cline{2-3}
& front $(f)$ & front right $(fr)$ \\
  \hline
  $\Gamma_{front}$ & $1.50\cdot10^6$ & $8.31\cdot10^5$ \\
  $\Gamma_{right}$ & $4.25\cdot10^5$ & $1.39\cdot10^6$ \\
  $\Gamma_{left}$ & $0$ & $0$ \\
  $\Gamma_{rear}$ & $0$ & $0$ \\
  $\Gamma_{host\ vehicle}$ & $1.93\cdot10^6$ & $2.22\cdot10^6$ \\
	\hline
\end{tabular}
\end{center}
\vspace{-2mm}
\end{table}
The resulting histograms of the collision probability rates are shown in fig. \ref{fig_Monte_Carlo_Histogram} together with the entry intensity obtained by numerical integration of the bivariate Gaussian in \Ref{eq_front_boundary_entry_intensity_cond} as well as the difference between the simulation and the calculation. The difference is calculated by evaluating the entry intensity at the same time as the mid points of the histogram bins.
As can be seen in fig. \ref{fig_Monte_Carlo_Histogram}, the entry intensity obtained by numerical integration of the exact expression (eq. \Ref{eq_front_boundary_entry_intensity_cond}) accurately reproduces the collision probability rate from Monte-Carlo simulations.

\begin{figure*}[tb!]
\centering
\null\hfill
\subfloat[Simulation based on $1\cdot10^5$ trajectories.]{\includegraphics[width = .6 \columnwidth]{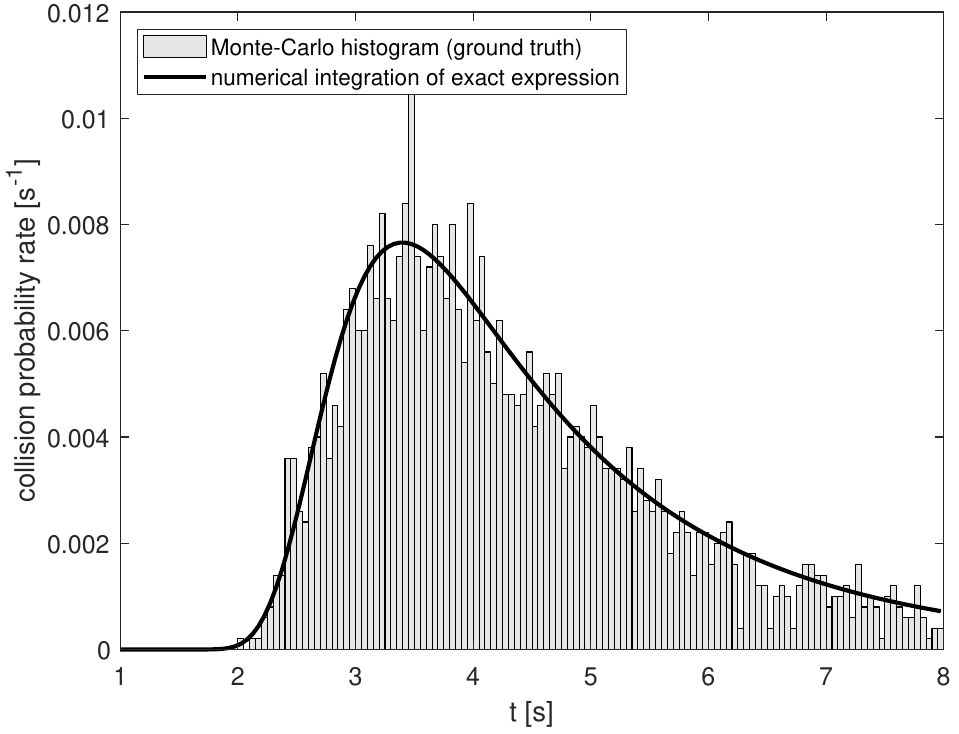}}
\hfill
\subfloat[Simulation based on $1\cdot10^6$ trajectories.]{\includegraphics[width = .6 \columnwidth]{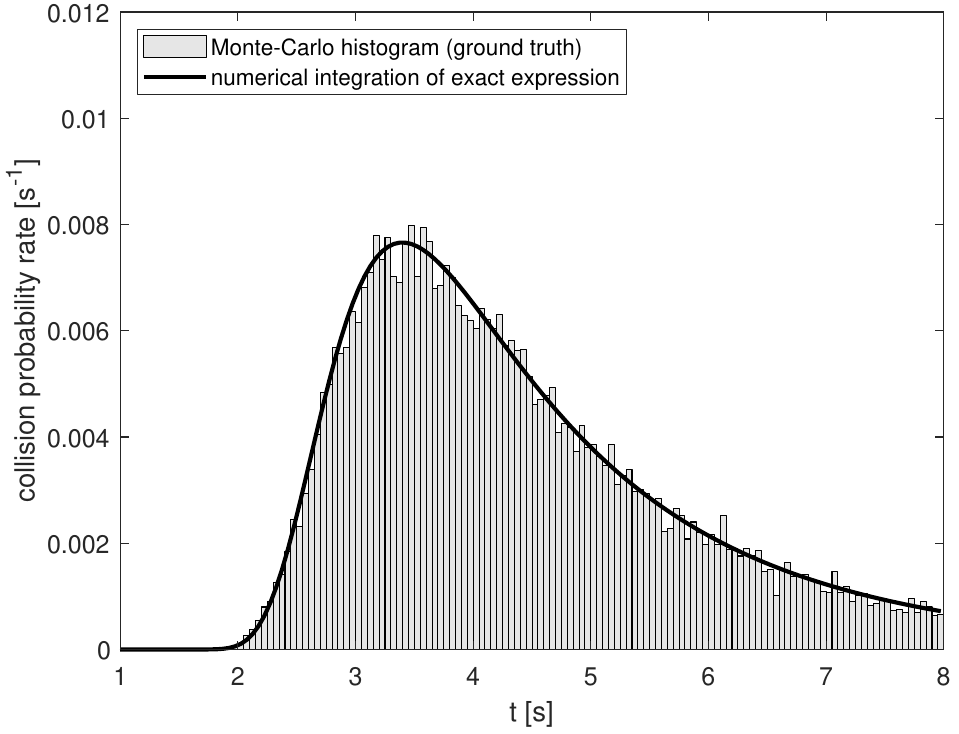}} 
\hfill
\subfloat[Simulation based on $1\cdot10^7$ trajectories.]{\includegraphics[width = .6 \columnwidth]{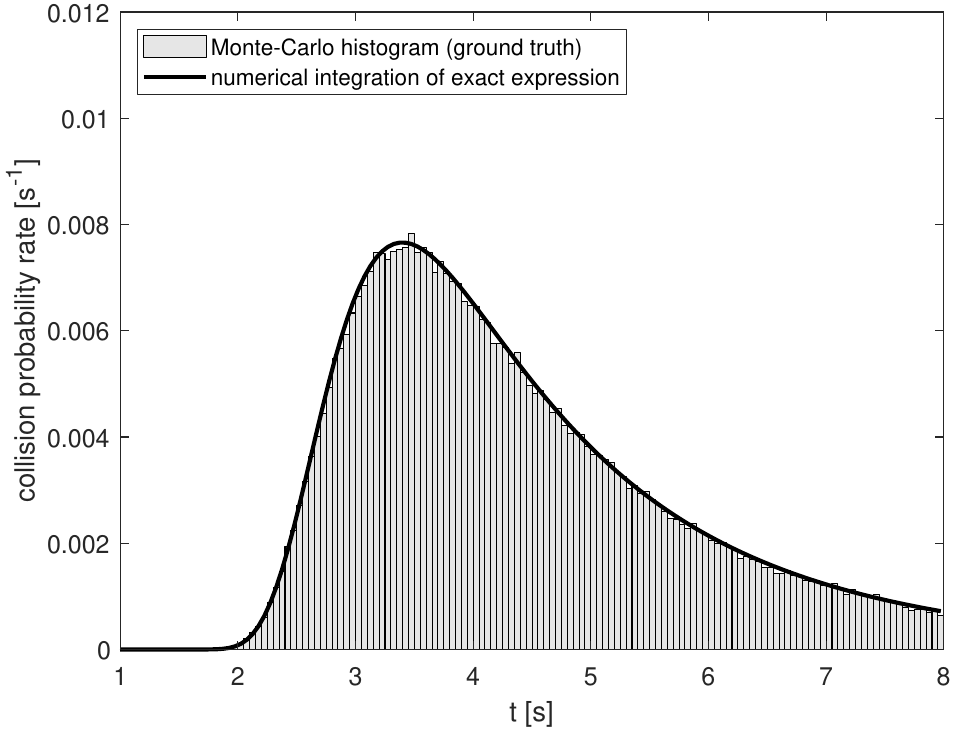}}
\hfill\null
\caption{The collision probability rate for the right side of the host vehicle for the front scenario is shown comparing the results from Monte-Carlo simulation with increasing amount of simulated trajectories (a)-(c) with the entry intensity obtained by numerical integration of the bivariate Gaussian distribution (eq. \Ref{eq_front_boundary_entry_intensity_cond}). The process noise PSD for both coordinates is $\tilde q_x = \tilde q_y = 1.0125 m^2 s^{-5}$.}
\label{fig_Monte_Carlo_Histogram_front_right_side}
\end{figure*}

\begin{figure*}[ht]
\centering
\null\hfill
\subfloat[Observed simulated trajectories\label{fig_trajectories_one_and_two_entries}]{\includegraphics[width = .9 \columnwidth]{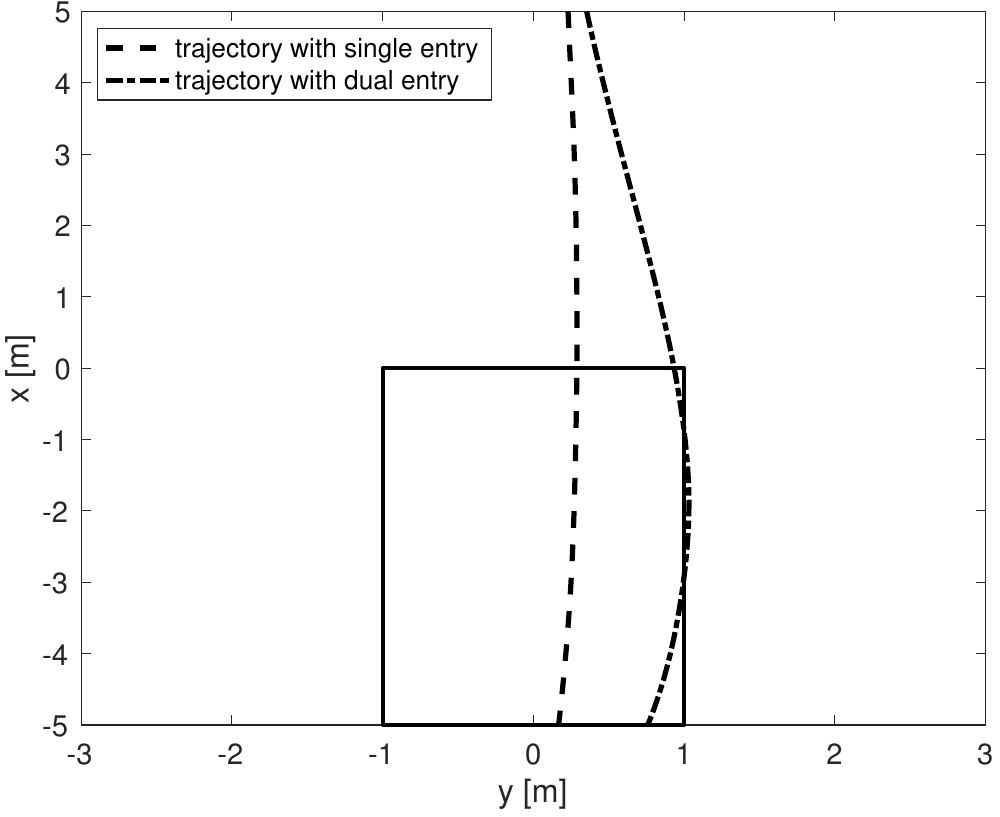}}
\hfill
\subfloat[Single and multiple entry histograms\label{fig_histogram_both_entries}]{\includegraphics[width = .9 \columnwidth]{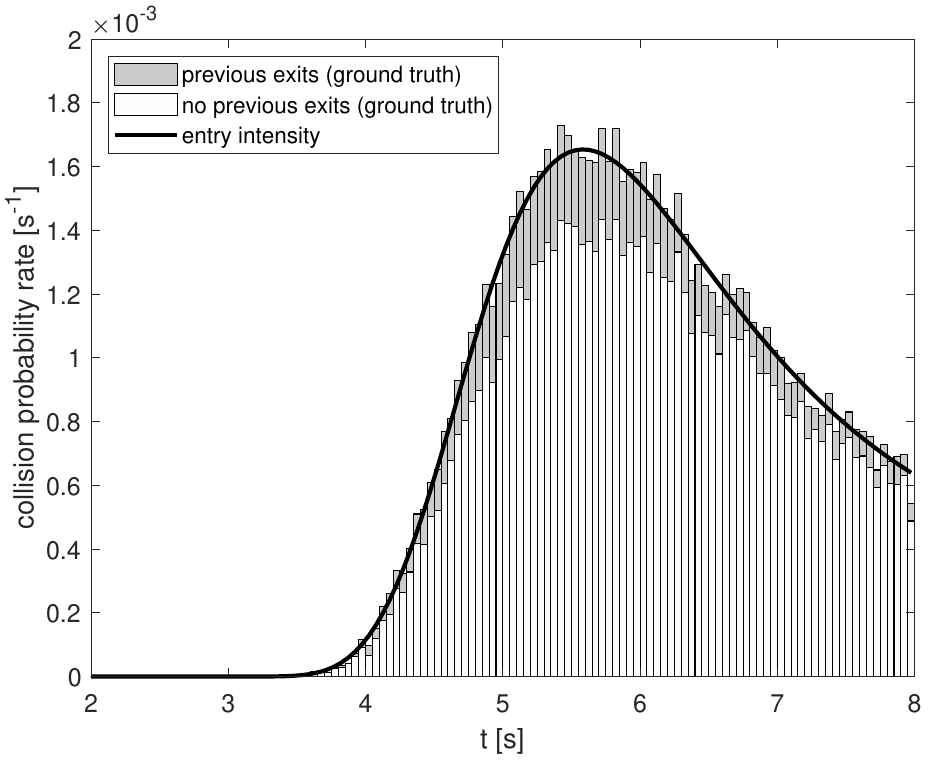}}
\hfill\null
\caption{Illustration of multiple entries into $\Gamma_{host\ vehicle}$. In (a), observed simulated a trajectory entering the entire vehicle boundary $\Gamma_{host\ vehicle}$ once and a trajectory entering twice are shown. The entry intensity of the right side $\Gamma_{right}$ of the host vehicle for the front scenario is shown in (b) together with the Monte-Carlo histogram where entries by trajectories that have previously exited $\Gamma_{right}$ from inside $\Gamma_{host\ vehicle}$ are marked in dark gray. Trajectories with no previous exit from inside are marked in light gray.}
\label{fig_multiple_entries}
\end{figure*}

In order to illustrate the increase in accuracy as a function of the number of simulated trajectories, fig. \ref{fig_Monte_Carlo_Histogram_front_right_side} shows the differences between simulation and numerical integration with increasing amount of simulated trajectories for collisions at the right side of the host vehicle in the front scenario. 
The reason why the entry intensity approximates the observed collision probability rates so well is the very low occurrence of higher order entries, i. e. entries where the trajectory enters the boundary more than once (see statistics of a Monte-Carlo simulation in table \ref{tab_coll_prob_full_boundary}). In the absence of higher order entries the expected number of entries becomes equal to the probability of entering the boundary at least once, see eq. \Ref{eq_expected_number_inequality}. Since the corresponding time interval is arbitrary this equality propagates to an equality of the rates (compare to eq. \Ref{eq_collision_probability_rate_upper_bound}). In this context, we want to point out a subtlety concerning the number of entries regarding the entire vehicle boundary $\Gamma_{host\ vehicle}$ versus entries through one of the boundary segments such as $\Gamma_{right}$. In Monte-Carlo simulations we have observed trajectories as shown in fig. \ref{fig_trajectories_one_and_two_entries} where the trajectory first enters the front boundary, exits the right boundary and then enters the right boundary again. With respect to the entire vehicle boundary $\Gamma_{host\ vehicle}$ this is a second entry -- however with respect to the individual right boundary segment $\Gamma_{right}$ this is a first entry. This is illustrated in fig. \ref{fig_histogram_both_entries} where the entry intensity and Monte-Carlo histogram for $\Gamma_{right}$ are plotted. Only by taking into account all entries for $\Gamma_{right}$, i. e. entries of $\Gamma_{right}$ that are first crossings of $\Gamma_{right}$, as well as entries of $\Gamma_{right}$ that are second or higher crossings does the entry intensity for $\Gamma_{right}$ match the histogram from Monte-Carlo simulation.
\begin{table}[h!]
\caption{Absolute frequency $H$ and relative frequency $P$ of the number of entries $N^+$ of colliding trajectories for $\Gamma_{host\ vehicle}$ based on $1\cdot10^7$ simulated trajectories for $\Delta T = [0,8s]$.}
\begin{center}
\label{tab_coll_prob_full_boundary}
\begin{tabular}{ m{1.5cm} m{1.5cm} m{1.5cm} m{1.5cm} }
 \hline
$X$ & $H(X)$ & $P(X)$ & $\frac{P(X)}{P(N^+\geq 1)}$ \\
  \hline
  $N^+\!\!= 1$ & $4,493,419$ & $0.4493$ & $0.9981$ \\
  $N^+\!\!= 2$ &     $8,772$ & $0.0009$ & $0.0019$ \\
  $N^+\geq 1$  & $4,502,191$ & $0.4502$ & $1$ \\
  \hline
\end{tabular}
\end{center}
\end{table}
\subsection{Does the approximation by Taylor-expansion accurately reproduce the exact result?}
\label{sec_numerical_study_Taylor}
\begin{figure*}[ht]
\centering
\null\hfill
\subfloat[Front scenario]{\includegraphics[width = .7 \columnwidth]{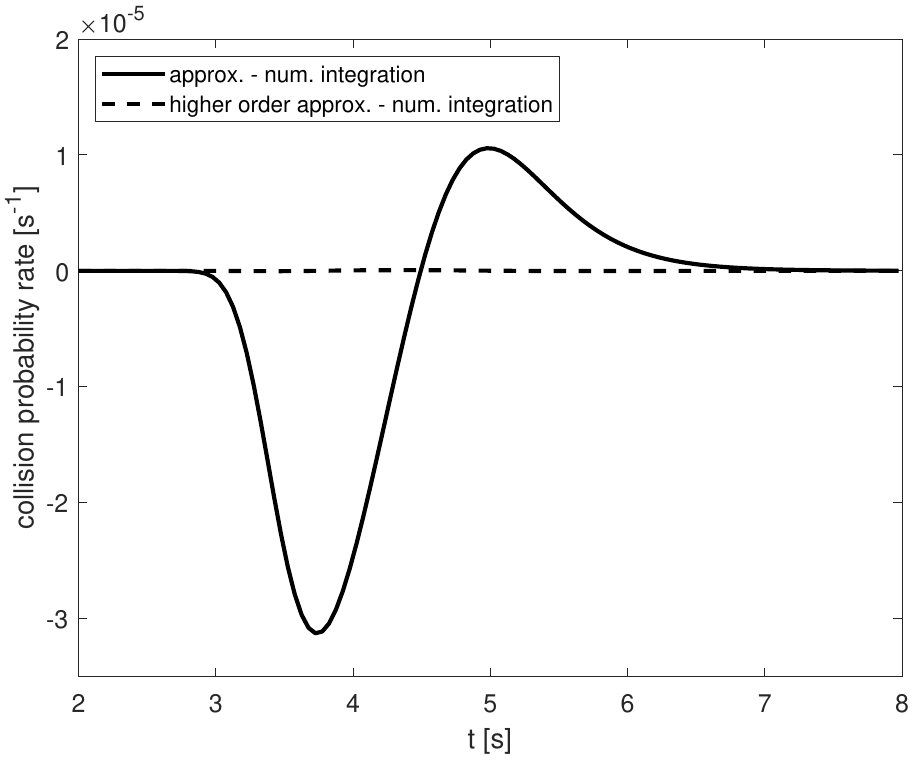}}
\hfill
\subfloat[Front-Right scenario]{\includegraphics[width = .7 \columnwidth]{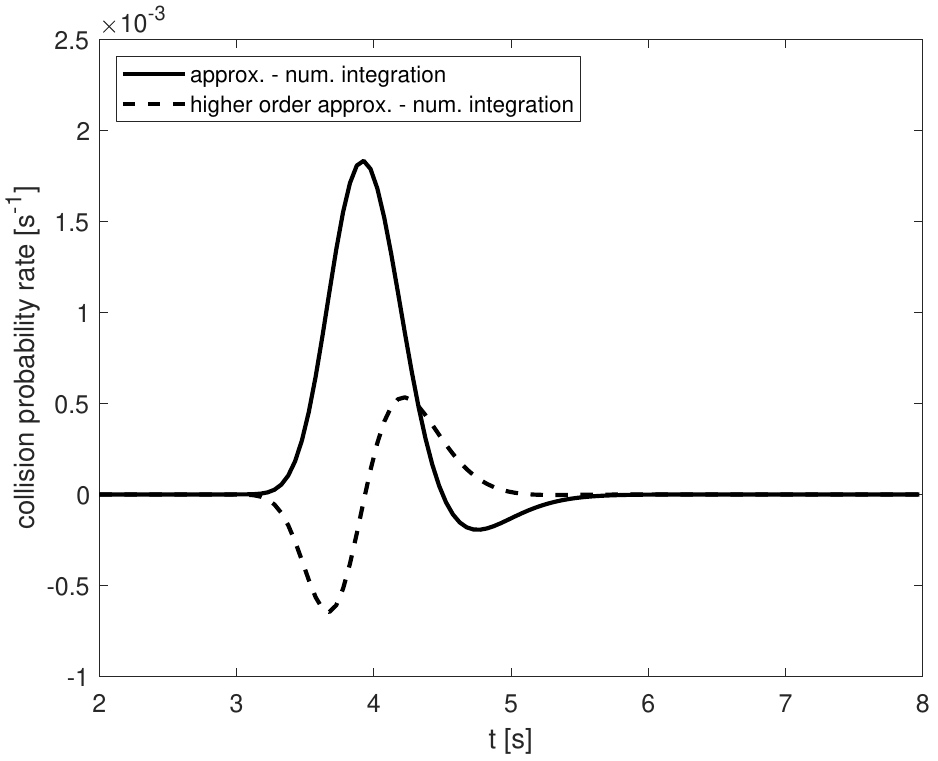}} 
\hfill\null\\
\null\hfill
\subfloat[Front scenario]{\includegraphics[width = .7 \columnwidth]{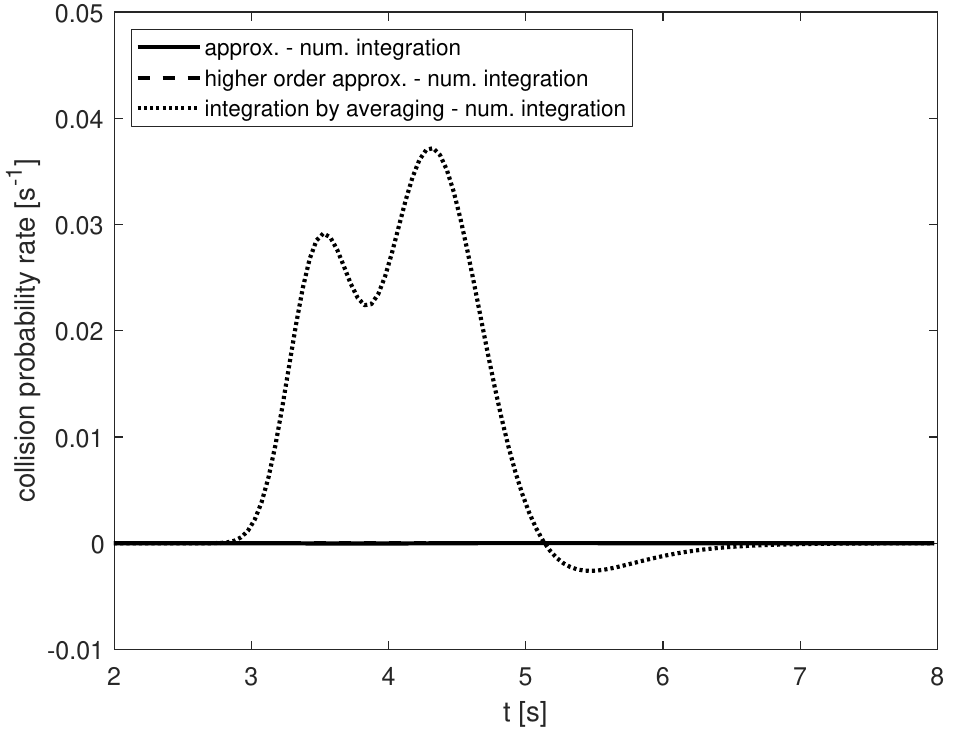}}
\hfill
\subfloat[Front-Right scenario]{\includegraphics[width = .7 \columnwidth]{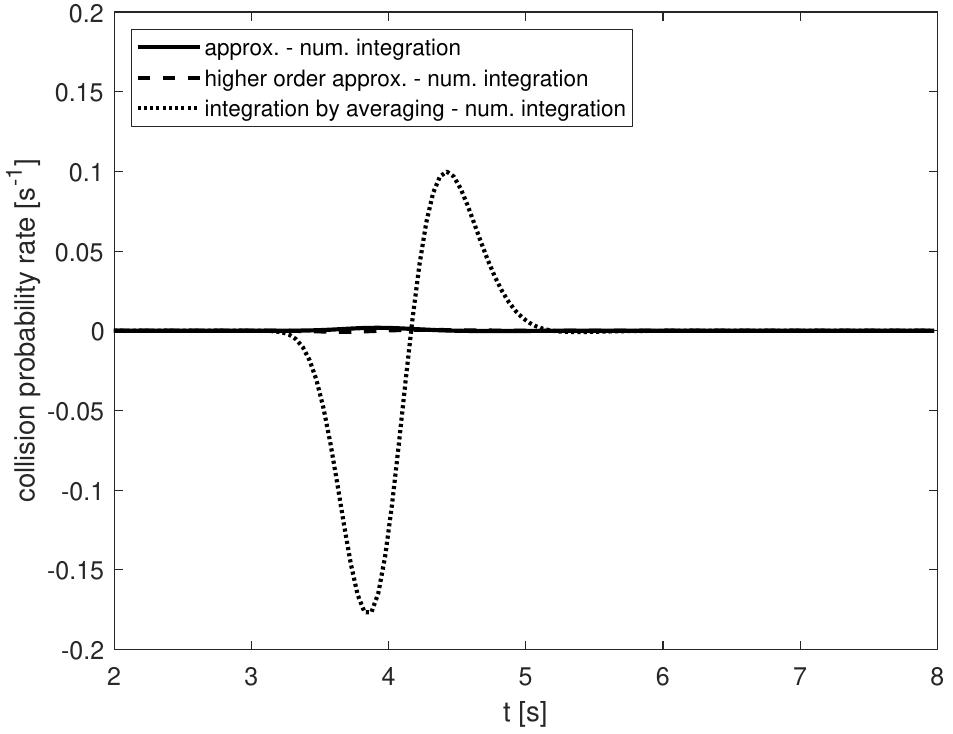}}
\hfill\null
\caption{Differences between numerical integration of the bivariate Gaussian in the expression of the entry intensity in eq. \Ref{eq_front_boundary_entry_intensity_cond} and two approximations (a,b) as well as the method described in \cite{blom2002conflict} (c,d).}
\label{fig_numerical_approximation_diff}
\end{figure*}

In order to be able to compute the entry intensity efficiently on an embedded platform, an approximation of the exact expression (eq. \Ref{eq_front_boundary_entry_intensity}) was derived in eq. (\ref{eq_coll_prob_rate_num_approx}). Fig. \ref{fig_numerical_approximation_diff}a,b shows the differences between this approximation as well as a higher-order approximation where the pdf is Taylor-expanded to linear order with respect to the off-diagonal element of the inverse covariance matrix around $0$ (see app. \ref{app_computation_integral}) and the numerical integration of \Ref{eq_front_boundary_entry_intensity_cond}. As can be seen, the higher-order approximation reduces the error to a large extent while it can be still calculated efficiently on an embedded platform using the complementary error function. In Fig. \ref{fig_numerical_approximation_diff}c,d the differences between the numerical integration of \Ref{eq_front_boundary_entry_intensity_cond} and the method described in \cite{blom2002conflict} is shown in addition where it can be seen that the deviation is much bigger compared to the approximations derived in (\ref{eq_coll_prob_rate_num_approx}).

\subsection{An adaptive method to sample the entry intensity over $\Delta T$}
\label{subsection_adaptive_method}
\begin{figure*}[!ht]
\centering
\null\hfill
\subfloat[Front scenario entry intensity]{\includegraphics[width = .7 \columnwidth]{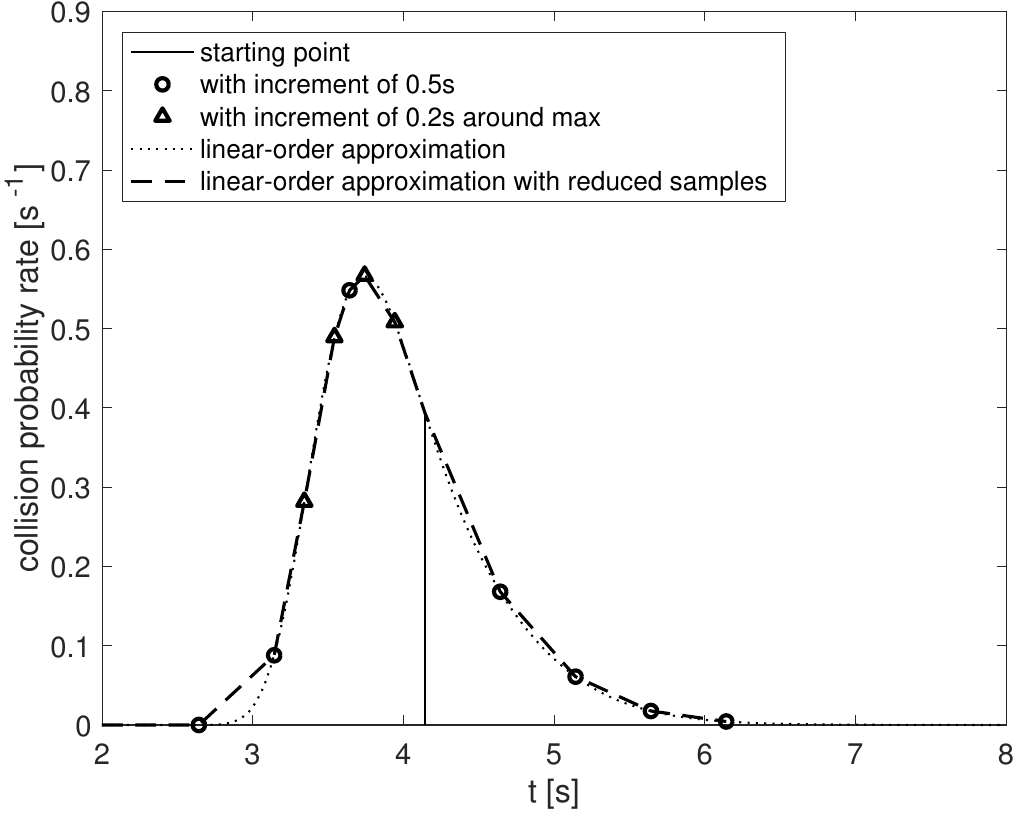}}
\hfill
\subfloat[Front-Right scenario entry intensity]{\includegraphics[width = .7 \columnwidth]{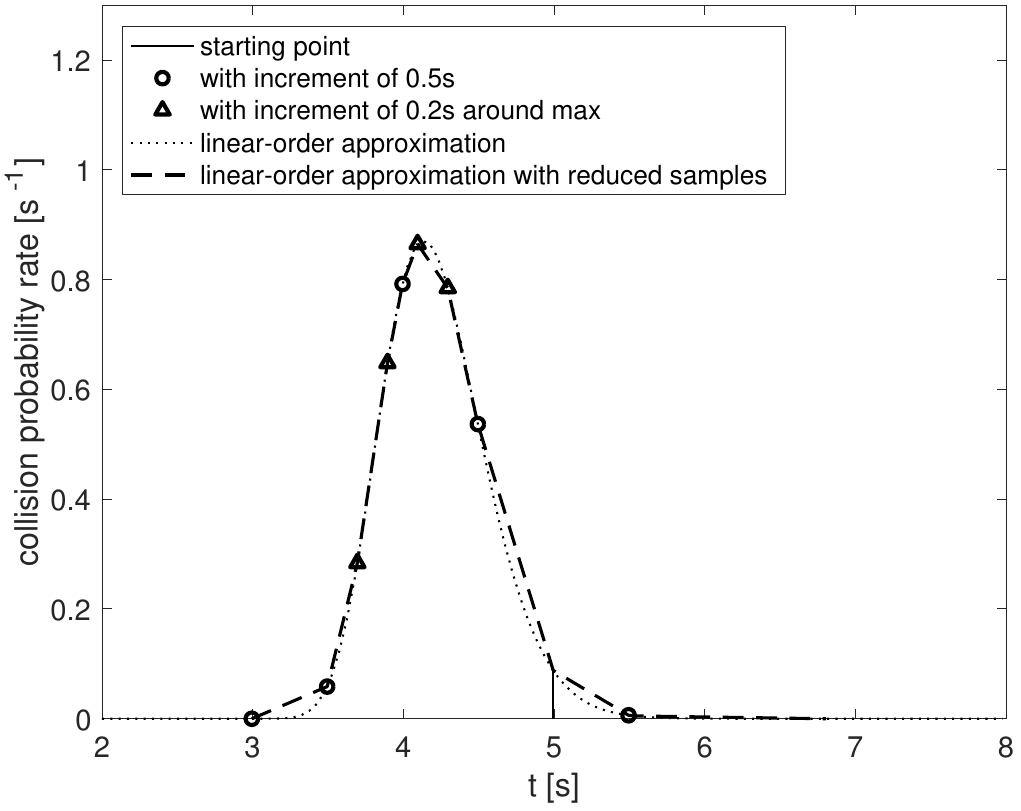}} 
\hfill\null\\
\null\hfill
\subfloat[Front scenario integrated entry intensity]{\includegraphics[width = .7 \columnwidth]{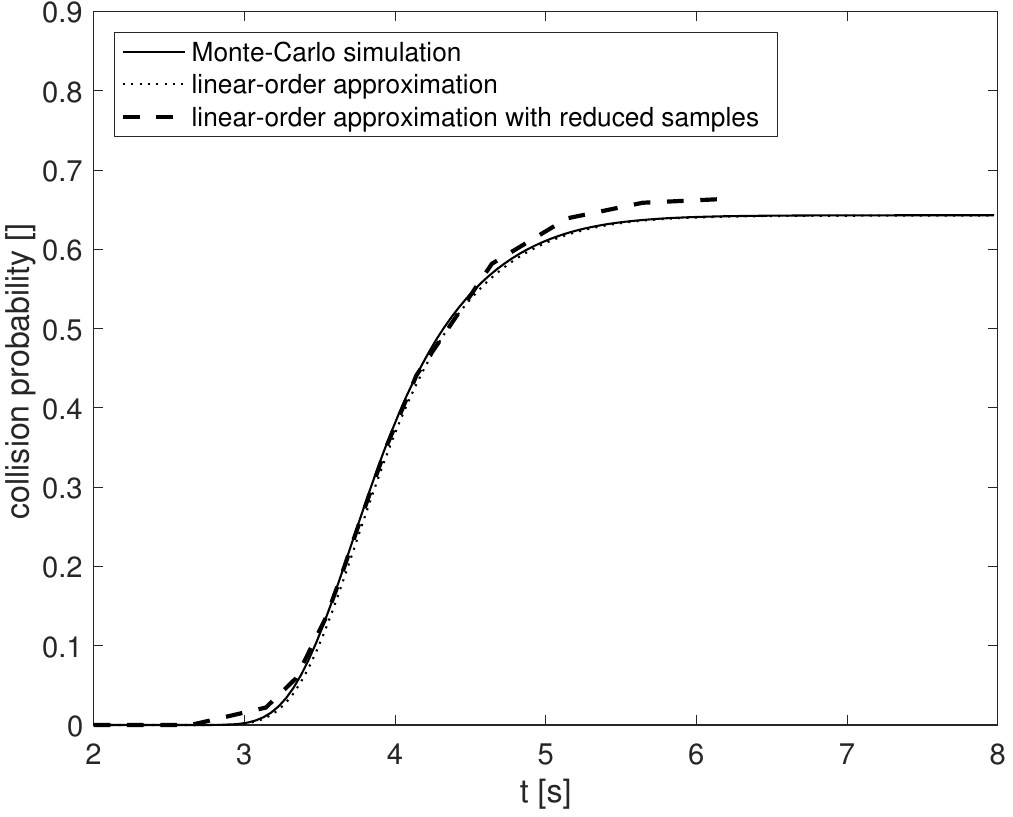}}
\hfill
\subfloat[Front-Right scenario integrated entry intensity]{\includegraphics[width = .7 \columnwidth]{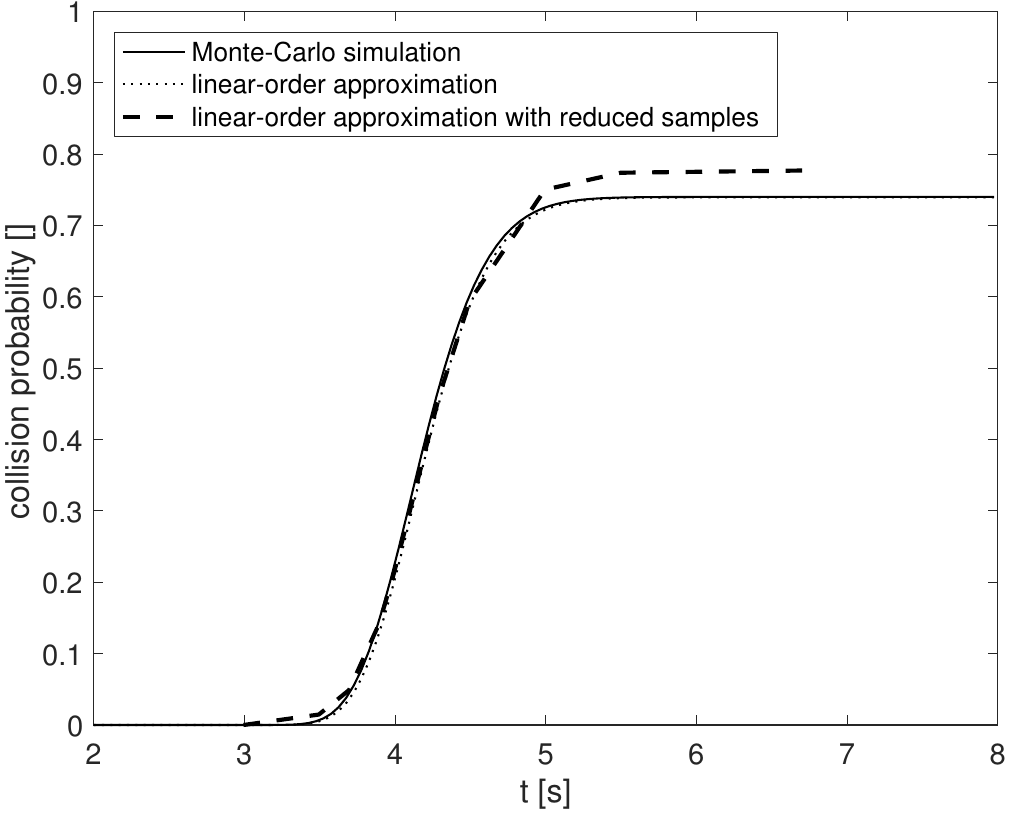}}
\hfill\null
\caption{Examples for reducing the number of calculations to determine the entry intensity and the integrated entry intensity. (a) and (c) show the results for the front scenario and (b) and (d) for the front-right scenario. The parameters in these examples are $\Delta t_1=0.5s$, $\Delta t_2=0.2s$ and $(dP_C^+/dt)_{low} = 0.01$. In doing so the number of calculations for the entry intensity could be reduced from $120$ (using a fixed sampling increment of $\Delta t = 0.05s$) to $13$ for the front scenario and to $12$ for the front-right scenario, respectively.}
\label{fig_delta_t_simulation}
\end{figure*}

The approximations above of the exact expression of the entry intensity were evaluated at small time increments of $\Delta t=0.05s$. Thus, the calculation over the entire time period of interest (e.g $8s$ as used above) and for every relevant object could induce a substantial computational burden. In order to reduce this effort, we propose an adaptive method to sample the entry intensity function with variable -- i. e. in general larger -- time increments $\Delta t$ over the time period of interest while still capturing the characteristics of this function, in particular its shape around the maximum.
The sampling starting point is based upon the non-probabilistic TTCs for single, straight boundaries using a one-dimensional constant acceleration model. 
Those TTCs for penetrating the front, left, and right boundaries can then be used as initial condition for the start of the sampling iteration of the entry intensity.\footnote{Due to the low probability of penetration the non-probabilistic TTC for the rear boundary is not considered for the determination of the sampling starting point.} To reproduce the entry intensity without substantial loss of information but with lower computational effort, the following algorithm is proposed: 
\begin{itemize}
\item Calculate the times of penetrating the front, left and right boundaries based upon the non-probabilistic TTCs described above.
\item Calculate the entry intensity for each time. Pick the time with the maximum entry intensity as a starting point.
\item Move left and right from this starting point with equally spaced $\Delta t_1 > \Delta t$ and calculate the entry intensity at these time points. Stop on each side if the entry intensity has reached a lower threshold of $(dP_C^+/dt)_{low}$. 
\item While moving left and right, check if the slope of the entry intensity has changed its sign.
\item On every slope sign change, calculate the entry intensity around this time interval with decreased $\Delta t_2 < \Delta t_1$.
\end{itemize}
Examples of this implementation can be found in fig. \ref{fig_delta_t_simulation} for the front and front-right scenarios. It can be seen that the entry intensity as well as the entry intensity integrated over a certain time period can be determined with considerably fewer sampling points while still capturing the shape of the functions to be approximated.

\subsection{Salient points of colliding vehicle's geometry}
\label{sec_salient_point_entry_intensities}
In this section, we investigate a family of entry intensities by a parsimonious sampling in terms of several representative salient points of the colliding vehicle's two-dimensional geometry, i.e. the four corner points of a vehicle's rectangular shape incorporating width and length information, see \ref{sec_entry_intensity_extended_vehicles_theory}. This enables the approximate estimation of the collision probability between two vehicles modeled as extended objects with arbitrary orientation in the horizontal plane by the collision probability of the ``riskiest" salient point.
\begin{figure*}[h!]
\centering
\null\hfill
\subfloat[Front left corner of colliding vehicle]{\includegraphics[width = .7 \columnwidth]{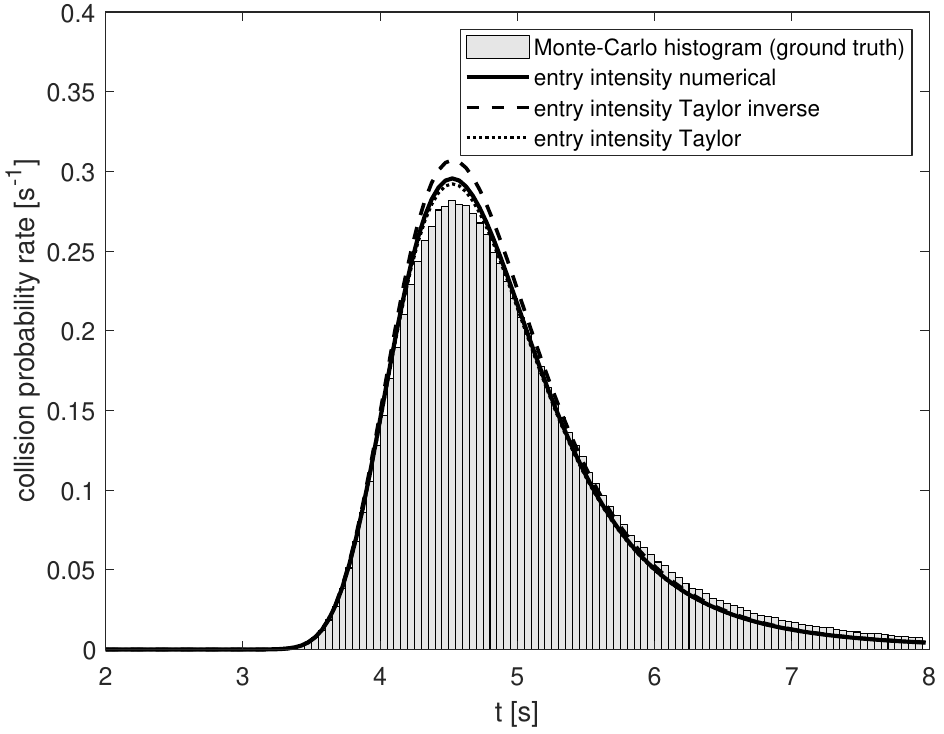}}
\hfill
\subfloat[Front right corner of colliding vehicle]{\includegraphics[width = .7 \columnwidth]{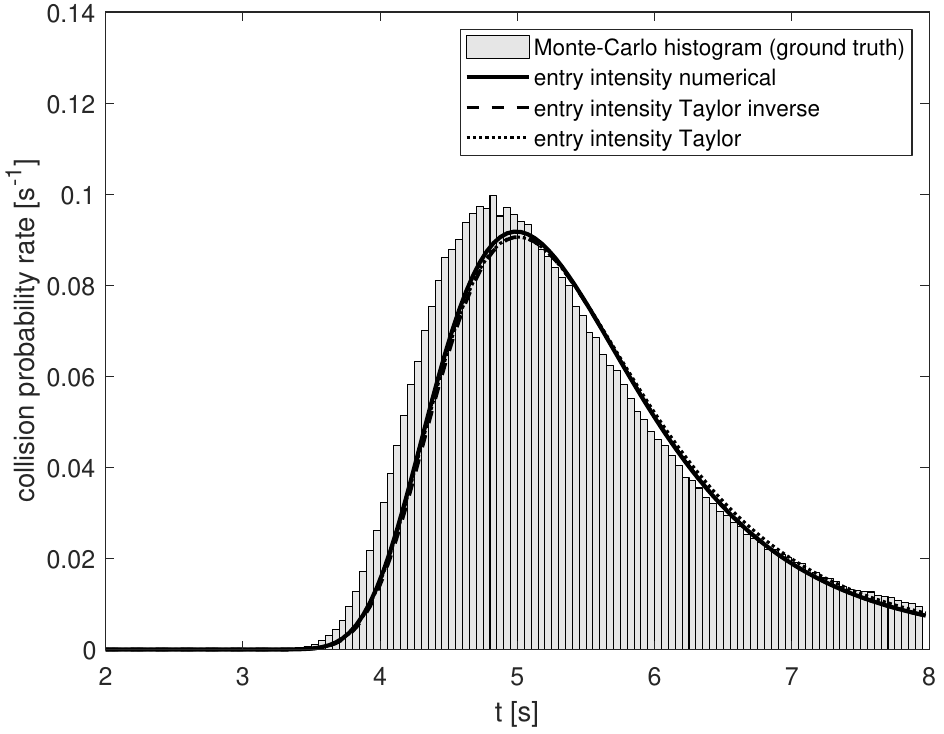}}
\hfill\null\\
\null\hfill
\subfloat[Rear left corner of colliding vehicle]{\includegraphics[width = .7 \columnwidth]{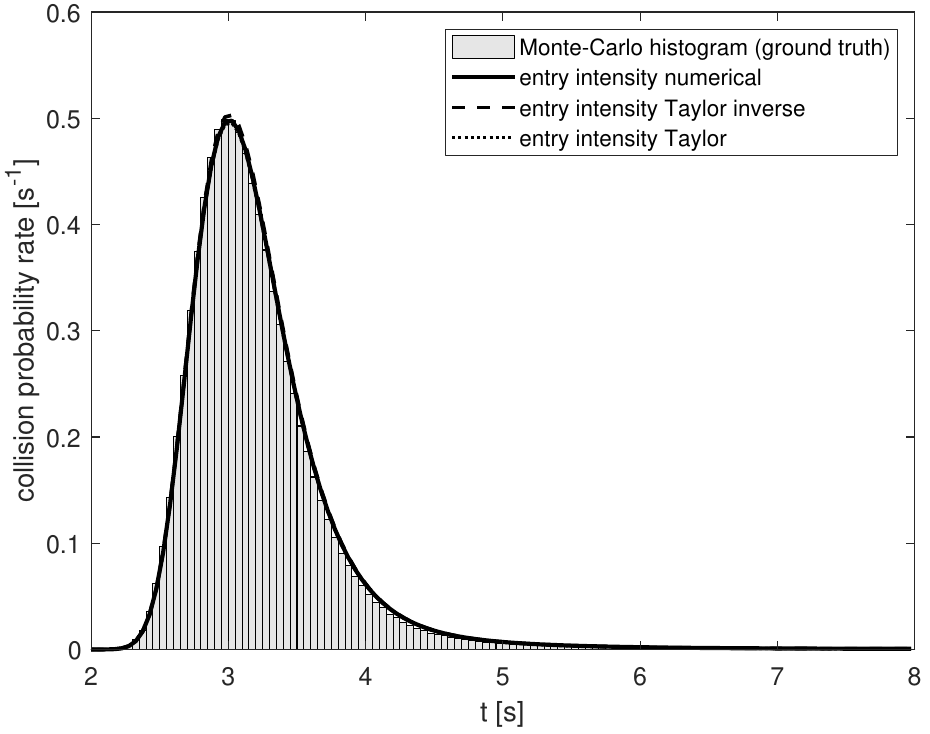}}
\hfill
\subfloat[Rear right corner of colliding vehicle]{\includegraphics[width = .7 \columnwidth]{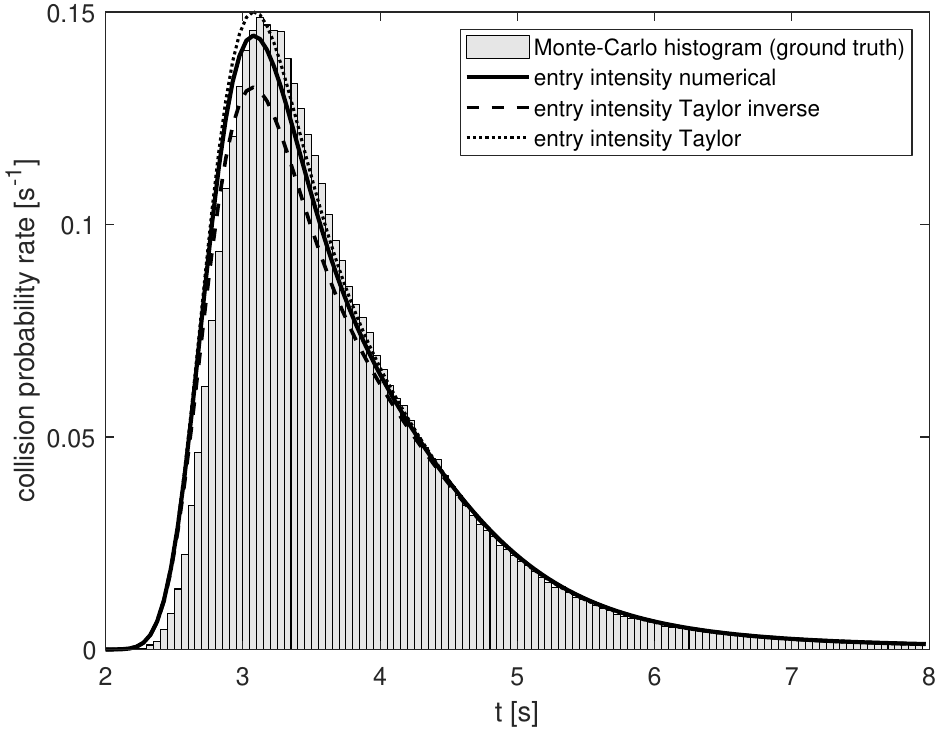}}
\hfill\null
\caption{Collision probability rate and entry intensity of four corner points in the front scenario with process noise PSD for both coordinates of $\tilde q_x = \tilde q_y = 0.0101 m^2 s^{-5}$ and input gain $B$ set to zero. Results for the entry intensity are given for numerical integration of the approximate 2d Gaussian distribution as well as two approximations to this integration as detailed in app. \ref{app_computation_integral}. It can be observed that the non-Gaussian nature of the probability distributions transformed to salient points entail
deviations with respect to Monte-Carlo simulations. This is due to the second order linearization with respect to the non-linear transformations derived in app. \ref{app_state_transformation_salient} for Gaussian densities.}
\label{fig_corner_points_of_colliding_vehicle}
\end{figure*}
The first step of the computation is the prediction of the reference point's state distribution to a certain time as before. But then it needs to be transformed to representative salient points as described in app. \ref{app_state_transformation_salient}. In order to apply the approximate formulae for Gaussian distributions as in sec. \ref{sec_entry_intensity_approx} the transformation is performed by the usual second order linearization, i. e. using the full nonlinear transformation for the mean and its Jacobian for the covariance matrix propagation.
For this investigation, three approaches are compared in fig. \ref{fig_corner_points_of_colliding_vehicle}: first the numerical integration of the resulting 2d Gaussian distribution as well as two closed-form approximations derived in app. \ref{app_computation_integral} by Taylor-expansion. Contrary to the investigations in sec. \ref{sec_numerical_corroboration} and \ref{sec_numerical_study_Taylor} even the numerical integration of the 2d Gaussian distribution cannot fully match the Monte-Carlo simulations due to the Gaussian approximation of the non-Gaussian transformed predicted distributions. Also both closed-form approximations to the 2d Gaussian integral show deviations to the Monte-Carlo simulation which describes the front scenario with process noise PSD for both coordinates of $\tilde q_x = \tilde q_y = 0.0101 m^2 s^{-5}$ and input gain $B$ set to zero. 
The closed-form approximations by Taylor-expansion with respect to the off-diagonal element of the covariance matrix and the inverse covariance matrix show similar accuracy with respect to the Monte-Carlo simulations except for the salient point in fig. \ref{fig_corner_points_of_colliding_vehicle}d where the former expansion is favored. Nevertheless in these cases both Taylor-expansions approximately capture both the shape and the location of the maximum of the intensity distributions.

\begin{figure}[!ht]
\centering
\includegraphics[width = .9 \columnwidth]{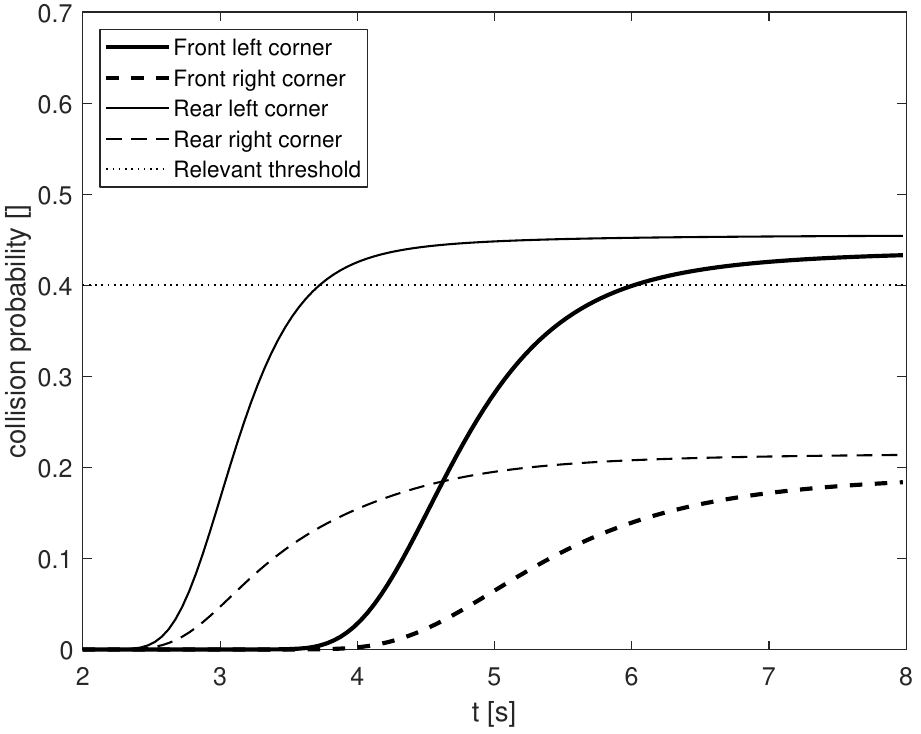}
\caption{Collision probability for the colliding vehicle's salient points. In this example the rear left corner is considered the ``riskiest" salient point, as it reaches a certain threshold the earliest.}
\label{fig_cumsum_corner_points_of_colliding_vehicle}
\end{figure}

The collision probability for the extended colliding object can then be approximated by the collision probability of the riskiest salient point which is the one where the collision probability exceeds a certain threshold the earliest. In the example above the riskiest salient point would be the rear left corner, see fig. \ref{fig_cumsum_corner_points_of_colliding_vehicle}. Clearly, using additional salient points such as the mid points of the vehicle's faces would improve the accuracy of this approximation at the expense of increased computational effort. Also, salient points can be used for more complicated, non-rectangular object boundaries.
In the next section we turn our attention to time-to-collision which is an often used characteristic of collision scenarios. 
\section{What is the TTC?}
\label{section_TTC}
The time-to-collision (TTC) is a stochastic process describing the first time of contact or collision and is hence characterized by a distribution. There have been various approaches to approximating the TTC-distribution. In \cite{Muntzinger_et_al_09}, the TTC is computed as the mean of the time distribution of reaching the $x_0$ boundary of the car as a function of the initial conditions assuming a constant speed model; process noise is not considered. This is also presented in \cite{jansson2008framework}; in addition the time distribution for reaching the $x_0$ boundary as a function of the initial conditions assuming a constant acceleration model is calculated by Monte-Carlo-simulation and its mean values depending upon the initial condition setup is given - again, process noise for this motion model is not considered. As a notable exception, in \cite{stellet2015uncertainty} the covariance of the distribution of TTC (or the related time-to-go in \cite{nordlund2008probabilistic}) has been augmented by standard error propagation and clever use of the implicit function theorem to include the effect of process noise. Nevertheless their TTC is still based on a reduction to a one-dimensional, longitudinal motion.
As will be shown below these restricted temporal quantities do not fully capture the characteristics of horizontal plane collision scenarios. What is required is a distribution of the TTC that takes into account process noise as well as two- or higher-dimensional geometries.

In the following figures entry intensities are plotted together with initial condition TTC-distributions from Monte-Carlo simulations similar to \cite{jansson2008framework}. These Monte-Carlo simulations are based on TTC values for the front boundary $\Gamma_{front}$ ($x$-direction) and the right boundary $\Gamma_{right}$ ($y$-direction) as solutions of the constant acceleration equations
\al{
x_{0} \alequal x(TTC_{front}) \nn\\
      \alequal x(0) + \dot x(0)\, TTC_{front} + { \ddot x(0) \over 2 } TTC_{front}^2 \nn\\
y_{R} \alequal y(TTC_{right}) \nn\\
      \alequal y(0) + \dot y(0)\, TTC_{right} + { \ddot y(0) \over 2 } TTC_{right}^2 \label{eq_deterministic_TTCs}
}
As an extension of the one-dimensional Monte-Carlo setup in \cite{jansson2008framework} the following conditions and constraints need to be considered for consistent TTC-histograms for one-dimensional boundaries embedded in two-dimensional space
\begin{itemize}
\item[-] for arbitrary initial conditions and values of $x_0, y_R$ all real, positive solutions of the quadratic equations above need to be considered
\item[-] a real, positive solution for $TTC_{front}$ is only valid if $\left( x(TTC_{front}), y(TTC_{front}) \right) \in \Gamma_{front}$, and a real solution for $TTC_{right}$ is only valid if $\left( x(TTC_{right}), y(TTC_{right}) \right) \in \Gamma_{right}$
\item[-] the trajectory must enter the boundary from outside, e. g. for $TTC_{right}$ it is checked that $y(TTC_{right} - \epsilon) > y_R$ for a small $\epsilon > 0$
\end{itemize}
Since time-dependent input cannot be handled in Monte-Carlo simulations only based on stochastic initial conditions we restrict the dynamical model in this section for comparison to a constant acceleration model, i. e. the input gain $B$ in app. \ref{app_vehicleModel} is set to zero.

\begin{figure}[ht]
\centering
\includegraphics[width = .9 \columnwidth]{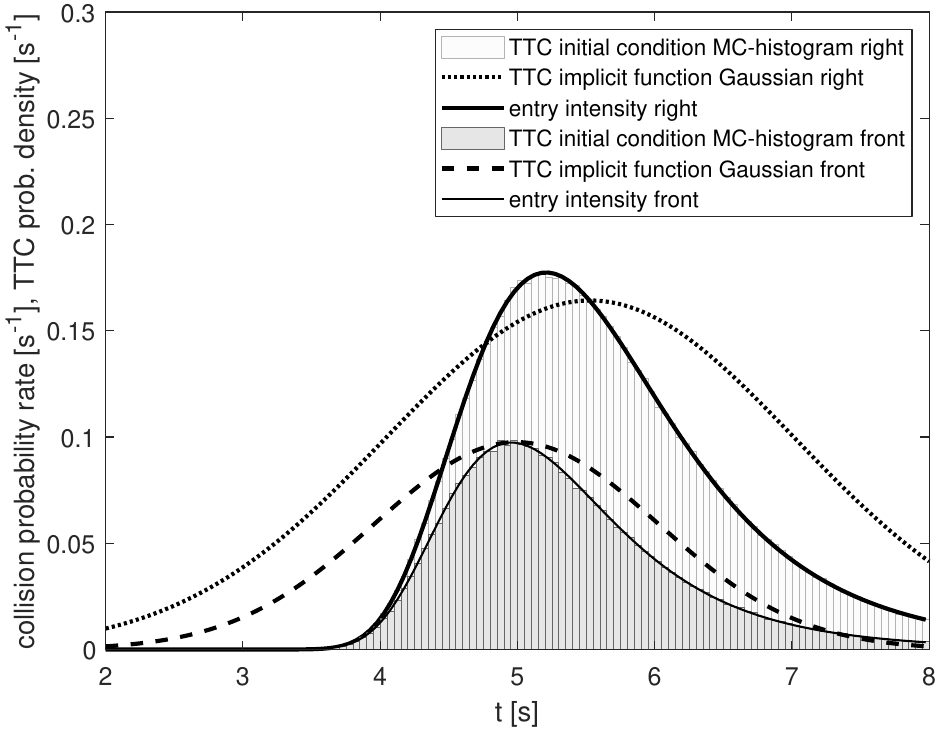}
\caption{Entry intensities, `TTC initial condition MC-histograms,' and `TTC implicit function Gaussians' for an initial condition at the front, right side of the vehicle: $(x,y)=(10,10)m$. For comparability, process noise had to be set to zero in the computation of the entry intensities.}
\label{fig_Monte_Carlo_TTC_versus_entry_intensity}
\end{figure}

As one central result of this section, we show in fig. \ref{fig_Monte_Carlo_TTC_versus_entry_intensity} that initial condition TTC-histograms from Monte-Carlo simulations described above (referred to as `TTC initial condition MC-histograms') match the corresponding entry intensities when process noise is zero. If contributions of higher order entries are negligible as discussed in sec. \ref{sec_numerical_corroboration} the entry intensity accurately approximates the collision probability rate which is the probability density of the TTC.
It is also noteworthy that in this case the entry intensity in its approximate version from sec. \ref{sec_entry_intensity_approx} affords a closed-form expression for a distribution that hitherto had to be obtained by Monte-Carlo simulation.

\begin{figure}[ht]
\centering
\includegraphics[width = .9 \columnwidth]{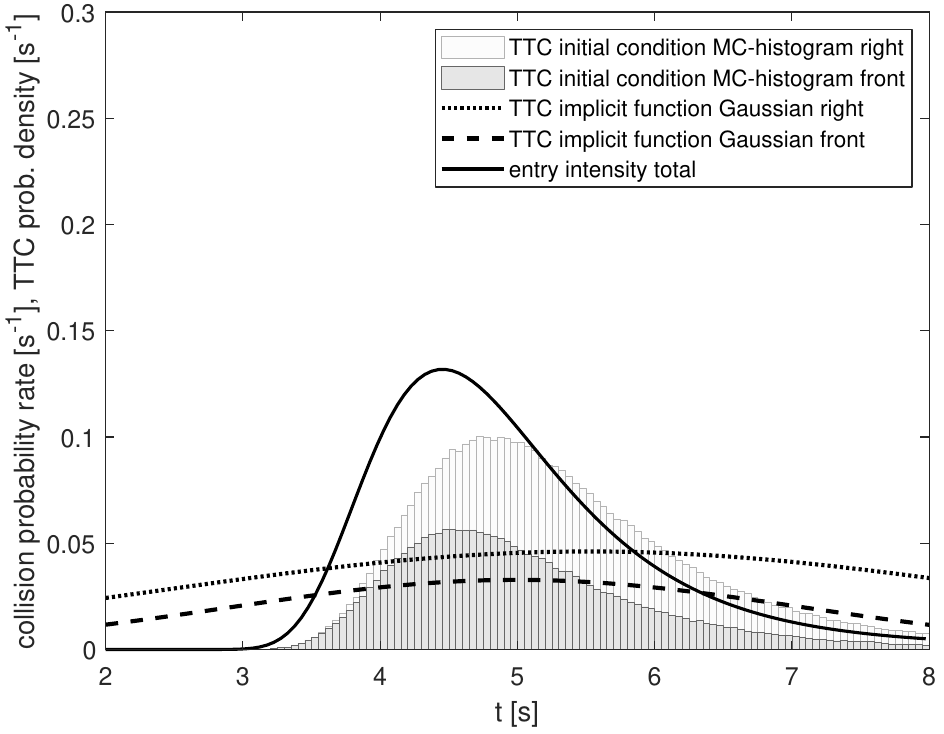}
\caption{Entry intensity, `TTC initial condition MC-histogram,' and `TTC implicit function Gaussian' for an initial condition at the front, right side of the vehicle: $(x,y)=(10,10)m$. The process noise PSD for both coordinates is $\tilde q_x = \tilde q_y = 0.0405 m^2 s^{-5}$.}
\label{fig_collision_probability_rate_and_TTCs}
\end{figure}

\begin{figure}[h]
\centering
\includegraphics[width = .9 \columnwidth]{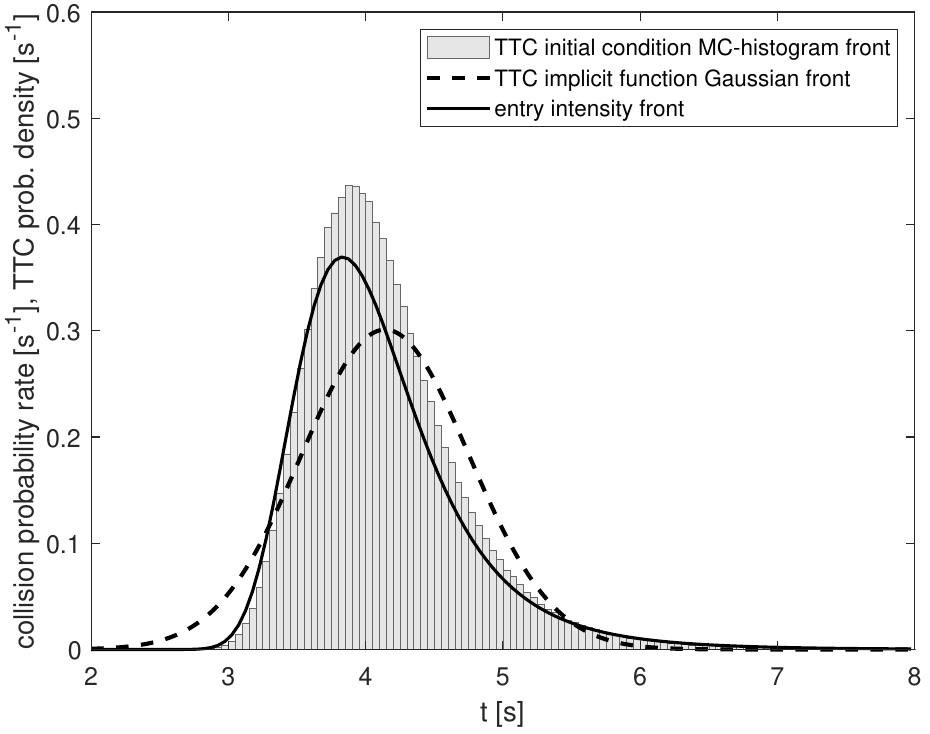}
\caption{Entry intensity, `TTC initial condition MC-histogram,' and `TTC implicit function Gaussian' for an initial condition in front of the vehicle: $(x,y)=(10,0)m$. The process noise PSD for both coordinates is $\tilde q_x = \tilde q_y = 0.0101 m^2 s^{-5}$.}
\label{fig_collision_probability_rate_and_TTCx_low_process}
\end{figure}

\begin{figure}[h]
\centering
\includegraphics[width = .9 \columnwidth]{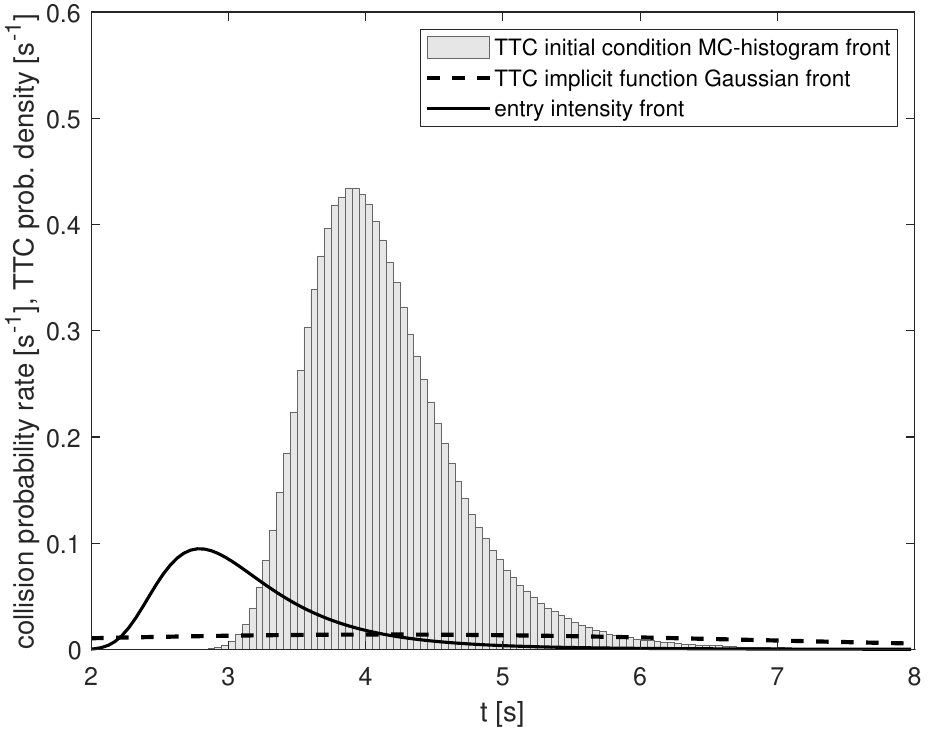}
\caption{Entry intensity, `TTC initial condition MC-histogram,' and `TTC implicit function Gaussian' for an initial condition in front of the vehicle: $(x,y)=(10,0)m$. The process noise PSD for both coordinates has been increased to $\tilde q_x = \tilde q_y = 1.0125 m^2 s^{-5}$.}
\label{fig_collision_probability_rate_and_TTCx_high_process}
\end{figure}

Also shown are the approximate Gaussian distributions according to the method using the implicit function theorem from \cite{stellet2015uncertainty} extended to a constant acceleration model to maintain comparability (referred to as `TTC implicit function Gaussians').
Their mean values coincide with the deterministic expressions of eq. \Ref{eq_deterministic_TTCs} due to the usual first-order approximation of non-Gaussian densities.\footnote{Note that the augmented TTC-computation in \cite{stellet2015uncertainty} does not alter the mean but only the covariance.}
Next, we turn on process noise and show how the `TTC initial condition MC-histograms' now deviate from the entry intensity.
In fig. \ref{fig_collision_probability_rate_and_TTCs} the entry intensity is plotted with `TTC initial condition MC-histograms' and `TTC implicit function Gaussians' for $x$- and $y$-directions for an initial position at the front, right side. 
The maxima and in particular the shapes of the `TTC implicit function Gaussians' in $x$- and $y$-direction are significantly different from the shapes and maxima of the entry intensity. Likewise, the `TTC initial condition MC-histograms' do not resemble the entry intensity and reach their maxima at later times.
Since the bulk of the colliding trajectories go through two sides - front and right (see also fig. \ref{fig_Monte_Carlo_sample_trajectory}b) - only a collision model that takes into account process noise and the full geometry of the host vehicle can yield accurate results.
In fig. \ref{fig_collision_probability_rate_and_TTCx_low_process} an entry intensity restricted to the $x$-direction is plotted together with a `TTC initial condition MC-histogram' and `TTC implicit function Gaussian.' The initial position is straight in front of the vehicle hence almost all trajectories pass through the front boundary. Nevertheless the entry intensity is lower and shifted to the left of the `TTC initial condition MC-histogram.' Also the maximum of the entry intensity occurs before the maxima of the `TTC initial condition MC-histogram' and the `TTC implicit function Gaussian.' In general the shapes and maxima
of the `TTC initial condition MC-histogram' and `TTC implicit function Gaussian' do not match the shapes and maxima of the entry intensity. 
These differences increase as the process noise increases as can be seen in fig. \ref{fig_collision_probability_rate_and_TTCx_high_process}. This is due to the fact that the time of the maximum is strongly influenced by the factor $p_t( x_0 )$ in eq. \Ref{eq_front_boundary_entry_intensity_cond}; an increased level of process noise leads to a faster spreading of $p_t( x_0 )$ and hence the maximum is reached earlier.
We have checked that the entry intensities in this section accurately match the ground truth histograms as in sec. \ref{sec_numerical_corroboration}.
The above discussion shows that temporal collision characteristics are encoded by the distribution of the entry intensity which incorporates the full geometry of the host vehicle as well as process noise during prediction.

A scalar quantity called {\it TTC} could then be obtained as one of the characteristic properties of this distribution such as the mode or the mean or the median, or as a property of the integrated collision probability rate, for example the time when the collision probability exceeds a certain threshold.  
\section{Conclusions}
As detailed in our literature review a common approach to compute a collision probability for automotive applications is via temporal collision measures such as time-to-collision or time-to-go. In this paper, however, we have pursued a different approach, namely the investigation of a collision probability rate without temporal collision measures as an intermediate or prerequisite quantity.
A collision probability rate then affords the provision of a collision probability over an extended period of time by temporal integration.
An expression for an upper bound of the collision probability rate has been derived based on the theory of level crossings for vector stochastic processes.
The condition under which the upper bound is saturated, i. e. is a good approximation of the collision probability rate has been discussed.
While the expression was exemplified by an application of Gaussian distributions on a two-dimensional rectangular surface, the formalism holds for general non-stationary as well as non-Gaussian stochastic processes and can be applied to any subsets of multidimensional piecewise-smooth surfaces.

The ground truth collision probability rate distribution has been obtained by Monte-Carlo simulations and approximated by our derived bound for the collision probability rate. We have also implemented an approximation of the collision probability rate bound that can be computed in closed form on an embedded platform. This approximate formula provided bounds of the collision probability rate distributions that are almost indistinguishable from distributions obtained by numerical integration for the scenarios considered in this paper. A straightforward application of this method characterizes the collision of an extended object with a second point-like object. The case of two extended objects with circular boundaries or rectangular boundaries with identical, fixed, axis-aligned orientation can also be reduced to the collision of an extended object with a second point-like object as illustrated in \cite{cir319}. In vehicle-to-vehicle collision scenarios these are unrealistic assumptions. Using an abstraction of the second object by salient points of its boundary we have shown how to augment the method to cover the case of two extended objects with arbitrary shape and orientation.

In our discussion of approaches to computing a TTC we illustrated the correspondence between classical TTC-distributions derived by Monte-Carlo simulations based on stochastic initial conditions and the entry intensity. We also showed that those classical one-dimensional TTC-distributions do not properly represent collision statistics in case of two-dimensional geometries and presence of process noise. The distribution of the collision probability rate is by construction the distribution of the TTC.

Point estimators derived from this distribution (e. g. the mode, mean, or median) or its temporal integral -- the collision probability -- could be investigated as input signals to collision avoidance decision making in the context of a complete collision avoidance system.
\section{ACKNOWLEDGEMENTS}
Helpful clarifications by Prof. Georg Lindgren are gratefully acknowledged.





\begin{appendix} 
\label{app_collision_probability}

\subsection{Partitioned Gaussian densities}
\label{app_Partitioned_Gaussian_densities}

In many calculations in stochastic estimation there is a need to marginalize over certain elements of a state vector or to obtain lower dimensional distributions by conditioning with respect to certain elements. For these calculations the original state vector $\xi$ can be rearranged or partitioned such that $x_r$ denotes the remaining state vector and $x_m$ denotes the states to be marginalized over or which are used for conditioning.
\eq{
\xi = \begin{pmatrix} x_r \cr x_m \end{pmatrix}
}
Hence the mean vector $\mu$ and covariance matrix $\Sigma$ can be partitioned into
\eq{
\mu = \begin{pmatrix} \mu_r \cr \mu_m \end{pmatrix},\quad \Sigma = \begin{pmatrix} \Sigma_{rr} & \Sigma_{rm} \cr 
\Sigma_{rm}^\top & \Sigma_{mm} \end{pmatrix}
}
The following two well-known results on multivariate Gaussians are used in this paper:

\paragraph{Marginalization}
The probability density of $\xi$ marginalized with respect to $x_m$ is
\eq{
p\left( x_r \right) = \int_{x_m} p\left( \xi \right) dx_m = \cN\left( x_r; \mu_r, \Sigma_{rr} \right)
}

\paragraph{Conditioning}
The probability density of $\xi$ conditioned on $x_m$ is
\al{
p\left( \xi | x_m \right) \alequal p\left( x_r | x_m \right) \nn\\                   
                          \alequal \cN\left( x_r; \mu_{r|m}, \Sigma_{r|m} \right)
}
with
\al{
\mu_{r|m}    \alequal \mu_r + \Sigma_{rm}\Sigma_{mm}^{-1}\left(x_m - \mu_m \right) \label{eq_conditional_mu}\\
\Sigma_{r|m} \alequal \Sigma_{rr} - \Sigma_{rm}\Sigma_{mm}^{-1}\Sigma_{rm}^\top \label{eq_conditional_cov_matrix}
}

\subsection{Dynamical system}
\label{app_vehicleModel}
The example kinematics is characterized by a six-dimensional state vector containing target vehicle coordinates relative to the host vehicle
\eq{
\xi = \begin{pmatrix} x & y & \dot x & \dot y & \ddot x & \ddot y  \end{pmatrix}^\top \label{eq_state_vector}
}
The continuous dynamics is given by a continuous white noise jerk model with additional time-dependent control input $u(t)$: 
\eq{
\dot\xi = F\xi + L\nu + B u  \label{eq_diff_equation}
}
where
\eq{
F = \begin{pmatrix}  0 & 0 & 1 & 0 & 0 & 0 \cr
                     0 & 0 & 0 & 1 & 0 & 0 \cr
                     0 & 0 & 0 & 0 & 1 & 0 \cr
                     0 & 0 & 0 & 0 & 0 & 1 \cr
									   0 & 0 & 0 & 0 & 0 & 0 \cr
									   0 & 0 & 0 & 0 & 0 & 0 \end{pmatrix},\quad
L = B = \begin{pmatrix}  0 & 0 \cr
												 0 & 0 \cr
												 0 & 0 \cr
												 0 & 0 \cr
												 1 & 0 \cr
												 0 & 1 \end{pmatrix} \nn
}
and 
\eq{
u(t) = \begin{pmatrix}  b_1 \sin( \omega t ) \cr
												b_2 \sin( \omega t ) \end{pmatrix} \nn
}
Process noise $\nu$ is characterized by the jerk power spectral density (PSD) $\tilde Q = \diag( \tilde q_x, \tilde q_y )$.

The discrete dynamics, i. e. the solution of this differential equation, can be obtained by
standard linear system techniques.
The covariance matrix of discrete-time equivalent process noise is given by (see e. g. \cite{BarShalomKirubarajan01}) 
\eq{
Q( t_{k+1}, t_k ) = \int\limits_{t_k}^{t_{k+1}}\Phi(t_{k+1}, \tau )L \tilde Q L^\top \Phi^\top(t_{k+1}, \tau ) d\tau \nn
}
where $\Phi$ is the transition matrix of the homogeneous differential equation.
The closed-form expression for this covariance matrix reads
\eq{
Q(\Delta t_k) = \begin{pmatrix} \frac{\Delta t_k^5}{20} & \frac{\Delta t_k^4}{8} & \frac{\Delta t_k^3}{6}\cr
\frac{\Delta t_k^4}{8} & \frac{\Delta t_k^3}{3} & \frac{\Delta t_k^2}{2}\cr
\frac{\Delta t_k^3}{6} & \frac{\Delta t_k^2}{2} & \Delta t_k\cr
\end{pmatrix} \otimes \begin{pmatrix} \tilde q_x & 0\cr
                                      0 & \tilde q_y \end{pmatrix}\nn
}
with $\Delta t_k = t_{k+1} - t_k$.
     
Note that the dynamical system above is an example to illustrate the application of the results in sec. \ref{sec_derivation_collision_prob} to compute the collision probability between two vehicles that can both be moving; other in general non-linear dynamical systems and state vectors can be used as long as they contain relative position and its first derivative. 

\subsection{Evaluation of the 2D integral for the entry intensity}
\label{app_computation_integral}

In this article integrals of the form 
\eq{
\int\limits_{\dot x \leq 0} \int\limits_{y \in I_y} \dot x\, p_t( y, \dot x | x_0 )\, dy d\dot x
}
as in eq. \Ref{eq_front_boundary_entry_intensity_cond} for the entry intensity appear. We are not aware of a closed-form solution if the covariance matrix of $p_t( y, \dot x | x_0 )$ is not diagonal.
In \cite{blom2002conflict} the 1D integral with respect to $\dot x$ was computed in closed form for a Gaussian pdf and the remaining spatial integral was replaced by the integrand at mid-point times the integration interval. As can be seen in figures \ref{fig_numerical_approximation_diff}(c,d) this approximation does not accurately reproduce the Monte-Carlo ground truth due to the considerable variation of the spatial distribution across the host vehicle rectangle. As an alternative approximation, we Taylor-expand the 2D pdf with respect to the off-diagonal element of the covariance matrix around 0 to a certain order and then integrate the factorized 1D distributions. For a general 2D Gaussian pdf $p( x_1, x_2 )=\cN(\xi;\mu,\Sigma)$ with
$\xi = ( x_1, x_2 )^\top$ and mean $\mu$ and covariance matrix $\Sigma$ the Taylor-expansion to linear order with respect to $\Sigma_{12}$ reads
\al{
\cN(\xi;\mu,\Sigma) \alequal \cN\left( x_1; \mu_1, \sqrt{\Sigma_{11}} \right) \cN\left( x_2; \mu_2, \sqrt{\Sigma_{22}} \right)\nn\\
              \alnothing + \Sigma_{12}\left( { x_1 - \mu_1 \over \Sigma_{11}} \cN\left( x_1; \mu_1, \sqrt{\Sigma_{11}} \right) \right)\cdot \nn\\
							\alnothing \qquad\cdot\left( { x_2 - \mu_2 \over \Sigma_{22}} \cN\left( x_2; \mu_2, \sqrt{\Sigma_{22}} \right) \right)\nn\\
							\alnothing + \cO\left((\Sigma_{12})^2\right) \nn
}
which leads to the following integral
\begin{multline*}
\int\limits_{x_{2l}}^{x_{2u}}\int\limits_{x_{1l}}^{x_{1u}} x_1 p( x_1, x_2 ) dx_1 dx_2 = \\
\left[ \mu_1 \Phi\left({ x_1 - \mu_1 \over \sqrt{\Sigma_{11}}}\right) - \Sigma_{11} \cN(x_1; \mu_1, \sqrt{\Sigma_{11}})
 \right]_{x_{1l}}^{x_{1u}}\cdot\\
\cdot\left[ \Phi\left({ x_2 - \mu_2 \over \sqrt{\Sigma_{22}}}\right)
      \right]_{x_{2l}}^{x_{2u}}\\
+ \Sigma_{12}\left[ \Phi\left({ x_1 - \mu_1 \over \sqrt{\Sigma_{11}}}\right) - x_1 \cN(x_1; \mu_1, \sqrt{\Sigma_{11}})
              \right]_{x_{1l}}^{x_{1u}}\cdot \\
\cdot\left[ - \cN(x_2; \mu_2, \sqrt{\Sigma_{22}})
      \right]_{x_{2l}}^{x_{2u}} + \cO\left((\Sigma_{12})^2\right)
\end{multline*}
The quality of the approximation depends asymptotically upon the size of $\Sigma_{12}$. An alternative Taylor-expansion would be an expansion with respect to the off-diagonal element of the {\it inverse} covariance matrix. Its off-diagonal element $\Sigma^{-1}_{12} := \left(\Sigma^{-1}\right)_{12} = -{\Sigma_{12}\over |\Sigma|}$ has the determinant of $\Sigma$ in the denominator, hence for large determinants (i. e. large uncertainties as expected for long prediction times) this approximation is expected to be more accurate. 
For a general 2D Gaussian pdf $p( x_1, x_2 ) = \cN(\xi;\mu,\Sigma)$ with
$\xi = ( x_1, x_2 )^\top$ and mean $\mu$ and covariance matrix $\Sigma$ the Taylor-expansion to linear order with respect to $\Sigma^{-1}_{12}$ reads
\al{
\cN(\xi;\mu,\Sigma) \alequal \cN\left( x_1; \mu_1, \sqrt{\tilde\Sigma_{11}} \right) \cN\left( x_2; \mu_2, \sqrt{\tilde\Sigma_{22}} \right)\nn\\
              \alnothing - \Sigma^{-1}_{12}\left( ( x_1 - \mu_1 ) \cN\left( x_1; \mu_1, \sqrt{\tilde\Sigma_{11}} \right) \right)\cdot \nn\\
							\alnothing \qquad\cdot\left( ( x_2 - \mu_2 ) \cN\left( x_2; \mu_2, \sqrt{\tilde\Sigma_{22}} \right) \right)\nn\\
							\alnothing + \cO\left((\Sigma^{-1}_{12})^2\right) \nn
}
with $\tilde\Sigma_{11} = {|\Sigma|\over \Sigma_{22}}, \tilde\Sigma_{22} = {|\Sigma|\over \Sigma_{11}}$.
This leads to the following integral 
\begin{multline*}
\int\limits_{x_{2l}}^{x_{2u}}\int\limits_{x_{1l}}^{x_{1u}} x_1 p( x_1, x_2 ) dx_1 dx_2 = \\
\left[ {\mu_1\over 2} \erf\left({ x_1 - \mu_1 \over \sqrt{2\tilde\Sigma_{11}}}\right) - \tilde\Sigma_{11} \cN\left(x_1; \mu_1, \sqrt{\tilde\Sigma_{11}}\right)  \right]_{x_{1l}}^{x_{1u}}\cdot\\
\cdot\left[ {1\over 2}\erf\left({ x_2 - \mu_2 \over \sqrt{2\tilde\Sigma_{22}}}\right) \right]_{x_{2l}}^{x_{2u}}\\
- \Sigma^{-1}_{12}\left[x_1 \tilde\Sigma_{11} \cN\left(x_1; \mu_1, \sqrt{\tilde\Sigma_{11}}\right) -{\tilde\Sigma_{11}\over 2}\erf\left({ x_1 - \mu_1 \over \sqrt{2\tilde\Sigma_{11}}}\right) \right]_{x_{1l}}^{x_{1u}}\cdot \\
\cdot\left[ \tilde\Sigma_{22}\cN\left(x_2; \mu_2, \sqrt{\tilde\Sigma_{22}}\right) \right]_{x_{2l}}^{x_{2u}} + \cO\left((\Sigma^{-1}_{12})^2\right)
\end{multline*}
If the covariance matrix of $p_t( y, \dot x | x_0 )$ is diagonal, i. e. $\Sigma_{12} = 0$, the integrand factorizes into Gaussians and can be integrated in a straightforward manner.

\subsection{State vector transformation to salient points}
\label{app_state_transformation_salient}
In order to transform the state distribution describing the object's reference point (such as the middle of the rear bumper or the middle of the rear axle) to other points such as the four corners the deterministic state transformation is needed, which can be used either by propagation of the mean and covariance using linear system techniques or by Monte-Carlo sampling. Transformation to other points of an extended object requires knowledge of its orientation which can be derived in the Ackermann limit from the angle of the velocity vector. This is an appropriate setup if the vehicle's reference point is the middle of the rear axle and side-slip at the rear wheels can be neglected as appropriate for normal driving conditions.
\begin{figure}[h]
\centering
\includegraphics[viewport = 7cm 5cm 20cm 13.2cm, clip, width = 0.6 \columnwidth]{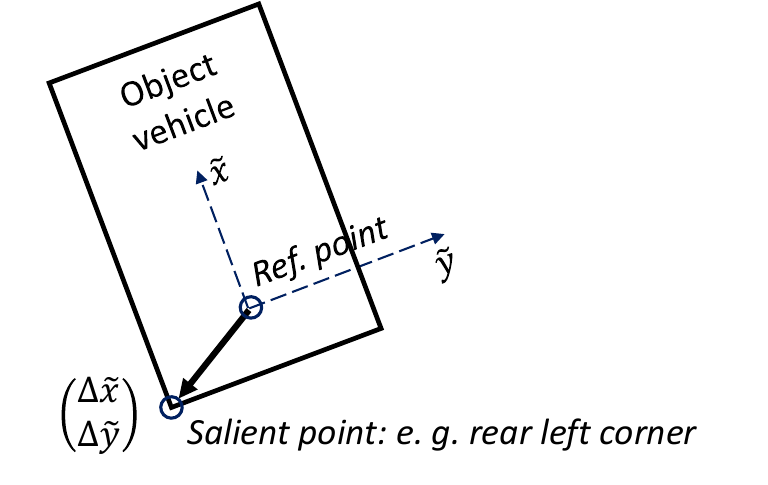}
\caption{Horizontal view of the object rectangle with local Cartesian coordinate system and coordinate origin at the
middle of the rear axle. The translation to the rear left corner as a salient point of the object's geometry is also drawn.}
\label{fig_object_rectangle_with_transformation}
\end{figure}
Taking into account the state vector as defined in eq. \Ref{eq_state_vector} and translating the state along $( \Delta \tilde x\  \Delta \tilde y )^\top$ in the object's local coordinate system (see fig. \ref{fig_object_rectangle_with_transformation}) the position transformation reads
\eq{
\begin{pmatrix} x \cr y\end{pmatrix}_{sal} = \begin{pmatrix} x \cr y\end{pmatrix}_{ref} + \bf{R} \begin{pmatrix} \Delta \tilde x \cr \Delta \tilde y\end{pmatrix} \nn
}
with
\eq{
\bf{R} = \begin{pmatrix} \cos\alpha & -\sin\alpha \cr \sin\alpha & \cos\alpha \end{pmatrix} \nn
}
and $\alpha = \arctan{\dot y \over \dot x}$ the orientation angle as explained above.
Then we have 
\al{
\begin{pmatrix} \dot x \cr \dot y\end{pmatrix}_{sal} \alequal \begin{pmatrix} \dot x \cr \dot y\end{pmatrix}_{ref} + \dot\alpha \bf{R}^\prime \begin{pmatrix} \Delta \tilde x \cr \Delta \tilde y\end{pmatrix} \nn\\
\begin{pmatrix} \ddot x \cr \ddot y\end{pmatrix}_{sal} \alequal \begin{pmatrix} \ddot x \cr \ddot y\end{pmatrix}_{ref} - {\dot\alpha}^2 \bf{R} \begin{pmatrix} \Delta \tilde x \cr \Delta \tilde y\end{pmatrix} + \ddot\alpha \bf{R}^\prime \begin{pmatrix} \Delta \tilde x \cr \Delta \tilde y\end{pmatrix} \nn
}
with $\bf{R}^\prime = {d\over d\alpha} \bf{R}$ and
\al{
\dot\alpha \alequal { \dot x \ddot y - \dot y \ddot x \over \dot x^2 + \dot y^2 } \nn\\
\ddot\alpha \alequal 2 { \dot x \dot y ( {\ddot x}^2 - {\ddot y}^2 ) - \ddot x \ddot y ( {\dot x}^2 - {\dot y}^2 ) \over (\dot x^2 + \dot y^2)^2 } + { \dot x \dddot y - \dot y \dddot x \over \dot x^2 + \dot y^2 } \nn 
}
Note that this transformation is non-linear, hence propagation of a multivariate Gaussian distribution by this transformation will result in a non-Gaussian distribution. In frameworks for Gaussian densities this can be handled by the usual second order linearization, i. e. using the full nonlinear transformation for the mean and its Jacobian for the covariance matrix propagation.

\end{appendix}


\bibliographystyle{IEEEtran}
\balance
\bibliography{bibliography}

\begin{thebibliography}{10}
\providecommand{\url}[1]{#1}
\csname url@samestyle\endcsname
\providecommand{\newblock}{\relax}
\providecommand{\bibinfo}[2]{#2}
\providecommand{\BIBentrySTDinterwordspacing}{\spaceskip=0pt\relax}
\providecommand{\BIBentryALTinterwordstretchfactor}{4}
\providecommand{\BIBentryALTinterwordspacing}{\spaceskip=\fontdimen2\font plus
\BIBentryALTinterwordstretchfactor\fontdimen3\font minus
  \fontdimen4\font\relax}
\providecommand{\BIBforeignlanguage}[2]{{%
\expandafter\ifx\csname l@#1\endcsname\relax
\typeout{** WARNING: IEEEtran.bst: No hyphenation pattern has been}%
\typeout{** loaded for the language `#1'. Using the pattern for}%
\typeout{** the default language instead.}%
\else
\language=\csname l@#1\endcsname
\fi
#2}}
\providecommand{\BIBdecl}{\relax}
\BIBdecl

\bibitem{jansson2008framework}
J.~Jansson and F.~Gustafsson, ``A framework and automotive application of
  collision avoidance decision making,'' \emph{Automatica}, vol.~44, no.~9, pp.
  2347--2351, 2008.

\bibitem{Muntzinger_et_al_09}
M.~M. Muntzinger, S.~Zuther, and K.~Dietmayer, ``Probability estimation for an
  automotive pre-crash application with short filter settling times,'' in
  \emph{Proceedings of {IEEE} Intelligent Vehicles Symposium}, 2009, pp.
  411--416.

\bibitem{stellet2015uncertainty}
J.~E. Stellet, J.~Schumacher, W.~Branz, and J.~M. Z{\"o}llner, ``Uncertainty
  propagation in criticality measures for driver assistance,'' in
  \emph{Intelligent Vehicles Symposium (IV), 2015 IEEE}.\hskip 1em plus 0.5em
  minus 0.4em\relax IEEE, 2015, pp. 1187--1194.

\bibitem{jansson2005collision}
J.~Jansson, ``Collision avoidance theory: With application to automotive
  collision mitigation,'' Ph.D. dissertation, Link{\"o}ping University
  Electronic Press, 2005.

\bibitem{lambert2008fast}
A.~Lambert, D.~Gruyer, and G.~Saint~Pierre, ``A fast {M}onte {C}arlo algorithm
  for collision probability estimation,'' in \emph{Control, Automation,
  Robotics and Vision, 2008. ICARCV 2008. 10th International Conference
  on}.\hskip 1em plus 0.5em minus 0.4em\relax IEEE, 2008, pp. 406--411.

\bibitem{nordlund2008probabilistic}
P.-J. Nordlund and F.~Gustafsson, \emph{Probabilistic conflict detection for
  piecewise straight paths}.\hskip 1em plus 0.5em minus 0.4em\relax
  Link{\"o}ping University Electronic Press, 2008.

\bibitem{mitici2018mathematical}
M.~Mitici and H.~A. Blom, ``Mathematical models for air traffic conflict and
  collision probability estimation,'' \emph{IEEE Transactions on Intelligent
  Transportation Systems}, vol.~20, no.~3, pp. 1052 -- 1068, 2019.

\bibitem{belyaev1968number}
Y.~K. Belyaev, ``On the number of exits across the boundary of a region by a
  vector stochastic process,'' \emph{Theory of Probability \& Its
  Applications}, vol.~13, no.~2, pp. 320--324, 1968.

\bibitem{bakker1993air}
G.~Bakker and H.~A. Blom, ``Air traffic collision risk modelling,'' in
  \emph{Decision and Control, 1993., Proceedings of the 32nd IEEE Conference
  on}.\hskip 1em plus 0.5em minus 0.4em\relax IEEE, 1993, pp. 1464--1469.

\bibitem{blom2002conflict}
H.~A. Blom and G.~Bakker, ``Conflict probability and incrossing probability in
  air traffic management,'' in \emph{Proceedings of the 41st IEEE Conference on
  Decision and Control}, vol.~3, 2002, pp. 2421--2426.

\bibitem{blom2003collision}
H.~Blom, B.~Bakker, M.~Everdij, and M.~Van Der~Park, ``Collision risk modeling
  of air traffic,'' in \emph{European Control Conference (ECC), 2003}.\hskip
  1em plus 0.5em minus 0.4em\relax IEEE, 2003, pp. 2236--2241.

\bibitem{cir319}
ICAO, ``A unified framework for collision risk modelling in support of the
  manual on airspace planning methodology for the determination of separation
  minima (doc. 9689), circ. 319-an/181,'' \emph{International Civil Aviation
  Organization, Montreal, Canada}, 2009.

\bibitem{prandini2000probabilistic}
M.~Prandini, J.~Hu, J.~Lygeros, and S.~Sastry, ``A probabilistic approach to
  aircraft conflict detection,'' \emph{IEEE Transactions on intelligent
  transportation systems}, vol.~1, no.~4, pp. 199--220, 2000.

\bibitem{genz2004numerical}
A.~Genz, ``Numerical computation of rectangular bivariate and trivariate normal
  and t probabilities,'' \emph{Statistics and Computing}, vol.~14, no.~3, pp.
  251--260, 2004.

\bibitem{belyaev1969characteristics}
Y.~K. Belyaev and V.~Nosko, ``Characteristics of excursions above a high level
  for a gaussian process and its envelope,'' \emph{Theory of Probability \& Its
  Applications}, vol.~14, no.~2, pp. 296--309, 1969.

\bibitem{lindgren1980model}
G.~Lindgren, ``Model processes in nonlinear prediction with application to
  detection and alarm,'' \emph{The Annals of Probability}, pp. 775--792, 1980.

\bibitem{lindgren2012stationary}
------, \emph{Stationary stochastic processes: theory and applications}.\hskip
  1em plus 0.5em minus 0.4em\relax CRC Press, 2012.

\bibitem{veneziano1977vector}
D.~Veneziano, C.~A. Cornell, and M.~Grigoriu, ``Vector-process models for
  system reliability,'' \emph{Journal of the Engineering Mechanics Division},
  vol. 103, no.~3, pp. 441--460, 1977.

\bibitem{illsley1998moments}
R.~Illsley, ``The moments of the number of exits from a simply connected
  region,'' \emph{Advances in Applied Probability}, vol.~30, no.~1, pp.
  167--180, 1998.

\bibitem{Altendorfer09}
R.~Altendorfer, ``Observable dynamics and coordinate systems for automotive
  target tracking,'' in \emph{Proceedings of {IEEE} Intelligent Vehicles (IV)
  Symposium}, 2009, pp. 741--746.

\bibitem{philipp2019analytic}
A.~Philipp and D.~Goehring, ``Analytic collision risk calculation for
  autonomous vehicle navigation,'' in \emph{2019 International Conference on
  Robotics and Automation (ICRA)}.\hskip 1em plus 0.5em minus 0.4em\relax IEEE,
  2019, pp. 1744--1750.

\bibitem{BarShalomKirubarajan01}
Y.~Bar-Shalom, X.~R. Li, and T.~Kirubarajan, \emph{Estimation with Applications
  to Tracking and Navigation}.\hskip 1em plus 0.5em minus 0.4em\relax Wiley,
  2001.

\end{thebibliography}

\end{document}